\documentclass[twocolumn]{aastex63}

\shorttitle{Close, Hyperbolic Encounters}
\shortauthors{Hansen }
 \watermark{}

\begin{document}

\title{ Unbound Close Stellar Encounters in the Solar Neighborhood}

\correspondingauthor{B. Hansen}
\email{hansen@astro.ucla.edu}

\author[0000-0001-7840-3502]{Bradley M. S. Hansen}
\affil{Mani. L. Bhaumik Institute for Theoretical Physics, Department of Physics and Astronomy,
University of California, Los Angeles, CA, 90095; hansen@astro.ucla.edu}




\begin{abstract}
We present a catalog of unbound stellar pairs, within 100~pc of the Sun, that are undergoing close, hyperbolic, encounters.
The data are drawn from the GAIA EDR3 catalogue, and the limiting factors are errors in the radial distance and unknown velocities along the
line of sight. Such stellar pairs have been suggested \citep{HZ21} to be possible events associated with the migration of technological civilisations between
stars. As such, this sample may represent a finite set of targets for a SETI search  based on this hypothesis.

Our catalog contains a total of  132 close passage events, featuring stars from across the entire main sequence, with  16 pairs featuring at least one main sequence star of spectral type between
K1 and F3. 
Many of these stars are also in binaries, so that we isolate  eight single stars as the most likely candidates to search for an
ongoing migration event --  HD~87978, HD~92577, HD~50669, HD~44006, HD~80790,  LSPM~J2126+5338, LSPM~J0646+1829 and  HD~192486.
 Amongst host stars of known planets, the stars GJ~433 and HR~858 are the best candidates.
\end{abstract}




\section{Introduction}

The accelerating pace of discovery of planets around other stars, including now the discovery
of potentially habitable planets, has brought into clear focus the question of how frequently
intelligent, technologically advanced, civilizations occur in the Galaxy. Searches for
signatures of extraterrestrial technology have a long history \citep{SETIhist}, but suffer from an enormous
needle-in-the-haystack problem, in that our capacity to search stars at a deep enough level
is limited to a very small subset of the available stars \citep{Tarter10,WKL18}.


Recently, we \citep{HZ21} -- hereafter HZ21 -- discussed the possibility that interstellar migration could be most
efficiently pursued during those brief episodes when unbound stars pass close to one another.
The energy costs of interstellar migration are prohibitive most of the time, but can be
substantially reduced during such encounters. Such considerations may become particularly
important for civilizations whose host stars are beginning to leave the main sequence.
Close encounters typically occur at intervals of
a few tens of million years (depending on exactly how close one wishes to get) and remain
optimal for a few thousand years. If a civilization were to avail itself of such an opportunity,
then the presence of two unbound stars in close proximity might serve as a signpost for a 
substantial enhancement in technological activity. A search for SETI signatures might then
have greater chance of success if directed towards such a sample.

The traditional approach of searching for extraterrestrial technology is to search in the radio, optical or infrared bands for evidence of
narrow-band signals characteristic of technological sophistication. A fundamental issue in searches of this kind is the
one of target selection, as the parameter space of potential sources is enormous \citep{Haystack}. Most searches are either broad in terms of source selection, such as \cite{HowHor,Isaac17,Enriquez17,TellisMarcy,Maire,Price20}, or focussed on systems
known to host exoplanets \citep{Siemion13,Margot18,SETI}. The proposal in HZ21 is based on a specific hypothesis and thus defines a 
much smaller, focussed, sample than other studies.
A search for technosignatures in these systems may also require some differences in search strategy, as traditional SETI searches assume a source located on a planet,
orbiting a star. In the case of a large-scale migration, and with sufficient angular resolution, sources may also be found to be localised between the stars in question, and
not solely in orbit around one star or the other.
 
Stellar interactions may also produce signals of astronomical interest even in the absence of technological effects. The
idea that close stellar passages sculpt the Oort cloud was an integral part of the original model of this population \citep{Oort50,Hills81}.
Although the tidal field of the galaxy is now held to be of greater importance in determining the properties of the Oort cloud \citep{HT86},
close stellar interactions are believed to be responsible for temporary enhancements in the rate of influx of comets
into the Solar system, so-called `comet showers' \citep{Hut87,FI87,Weiss96,Dyb02}.

In this paper we therefore seek to identify pairs of stars that meet  the criterion of being
in close physical proximity, but which are gravitationally unbound.
 These may
serve as a possible target list for  a SETI search -- one with  the particular motivating
hypothesis  laid out in HZ21. To do this, we draw on the results of the GAIA satellite, which
 has provided a detailed map of parallaxes and proper motions for nearby stars.  We
 search for pairs of stars that have a finite probability of passing within $10^4$~AU of one another -- either in the recent past or near future.
  This threshold is chosen to be such that these events are sufficiently uncommon (only $\sim 10^{-4}$ G stars will be involved in such an encounter at any one time)
 to justify the hypothesis in HZ21 that migration is limited to rare events. 
In section~\ref{Close} we describe the selection of a sample of stellar pairs that satisfy the above criterion,  and which lie within 100~pc of the
Sun.  The accuracy of our stellar matches decreases with distances and we choose the threshold of 100~pc to be large enough to yield a
 sample of reasonable size while mitigating the uncertainties.
 In section~\ref{Sigs} we 
discuss possible technological and astrophysical signatures associated with these events.

\section{Close but unbound pairs}
\label{Close}

The GAIA Early Data Release 3 -- EDR3 --\citep{GAIA0,EDR3,Lind3}, provides parallaxes and proper motions for 
several billion stars surrounding the Sun. With distances, relative angular separations, and relative velocities,
we may try to project the encounter histories of pairs of stars into the near future and recent past, to estimate
how close they have/will come. Several studies have investigated the sample of wide binaries identifiable in the
GAIA data \citep{HL20,Smart21,BadBin}. Our interest is in the systems these analyses reject, in that we wish to
identify systems that are in close physical proximity but which are not gravitationally bound to one another.

Our first cut is to select all pairs of stars which have
parallaxes whose naive conversion to distance places the radial distances within 0.5~pc of one another, and  whose projected separations on the 
sky lie within $3 \times 10^4$AU, where we use the
closest of the two parallaxes to convert angular separation into distance. This procedure is dictated
by the fact that the uncertainty in the radial distance is greater than the uncertainties in the
angular separation. We adopt this relatively loose cut  to define
a sample that we will refine further below. We will also restrict our sample to those stars within 100~pc
of the Sun. This is also dictated by the increase in error in the radial separation as the distance
increases. Finally, we use the GAIA astrometric error cut {\tt astrometric\_sigma5d\_max$<1$} to remove
those sources with  potentially problematic astrometry. Additional cuts based on photometric precision can
produce cleaner stellar samples (e.g. \cite{Babu18}), but we want our initial selection to err on the side of
inclusion, because our final candidate sample will be subject to additional quantitative vetting.

This first cut identifies stars that are currently in close physical proximity. Many of
these will be genuine bound binaries. The GAIA  catalogue also provides proper motions, which can be converted
into velocities. At the separations of interest, the orbital velocity of stellar mass binaries is much
less the expected magnitude of relative velocities of unbound disk stars, so we expect that
genuinely bound objects should exhibit common proper motions to a level that distinguishes
them from unbound pairs. The quantification
of proper motion errors has its own uncertainties, but these are small enough that one
can identify binary samples of high fidelity \citep{BadBin}. 

\begin{figure}
\includegraphics[height=6cm, width=5cm, angle=0, scale=1.75]{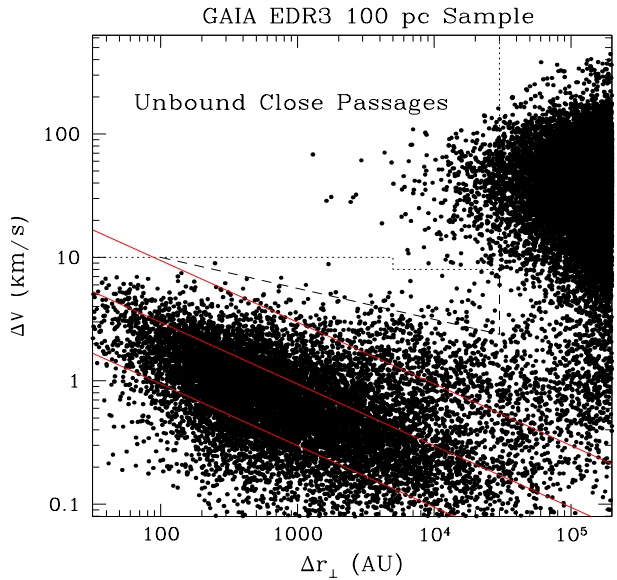}
\caption{ The points show all pairs of stars within 100~pc,
whose projected separations are less than $2 \times 10^5$~AU. The  solid red lines indicate
the escape velocity from an orbit of the stated $\Delta r_{\perp}$, for three different total
system masses -- $0.1 M_{\odot}$ (lowest curve), $1 M_{\odot}$ and $10 M_{\odot}$ (uppermost line).
 The dotted lines enclose the subset of   446 stellar pairs that we
select for further study.  The points between the dashed line and the dotted lines show a very high
incidence of unmodelled astrometric noise and are likely to be hierarchical triples.
 \label{Split}}
\end{figure}

Figure~\ref{Split} shows the  how the relative velocity of stars varies with their relative separation from their nearest neighbours
on the sky. We plot relative velocity versus
separation on the sky for all pairs whose distances are within 0.5~pc of each other and
 whose projected sky-plane separations are $\Delta r_{\perp}< 2 \times10^5$~AU.
 As noted by prior authors,  GAIA proper motions effectively separate binaries (towards the lower left of the diagram)
from the unbound pairs (towards the upper right of the diagram). The  three solid red lines indicate the
escape velocity from the corresponding orbital separation,  for binaries of total mass $0.1 M_{\odot}$, $1 M_{\odot}$ and $10 M_{\odot}$, from bottom to top. 
  The projected separation shown here is less than, or equal to, the true separation of the corresponding binary and so systems that lie below these curves have relative velocities consistent with orbital motion of bound pairs. The sample to the upper right represents the pairs of
stars that are moving too fast to be gravitationally bound to one another.
 This group has a small tail to low
projected separations, and it is in this group that we are interested. 

 There is also a substantial population of stars that lie above the uppermost red line. Although these appear to be contiguous to the 
bound sample, their interpretation would require system masses well in excess of $10 M_{\odot}$. Although this is technically possible, their
frequency is unrealistic within our understanding of stellar demographics and lifetimes. We have thus investigated this population in more detail.
In particular, we determine that the bulk of these pairs have relative motions that are contaminated by orbital motions of an unresolved binary in
at least one member of the pair. To support this, we downloaded, for each star, the GAIA {\tt ruwe} (renormalised unit weight error) parameter, which measures
the excess astrometric noise remaining after the astrometric fit of the position and motion of the source in question. A value of {\tt ruwe}$>1.4$ is considered
to be an indicator of unmodelled astrometric noise and a signature of a partially resolved binary.  For the sources that lie between the dashed
and dotted lines in Figure~\ref{Split}, we find that 56/67 (84\%) of these sources have {\tt ruwe}$>1.4$. We conclude that the bulk of the points at the
upper edge of the bound sample are hierarchical triples (at least), in which the orbital motion of an unresolved binary has resulted in a biased measurement
of the relative motion of the wider pair. We therefore adopt a criterion given by the dotted line in Figure~\ref{Split}. Our  initial  sample includes
pairs where  $\Delta r_{\perp}<3 \times 10^4$~AU and $\Delta v > 8$~km~s$^{-1}$, but with a higher threshold of $\Delta v>10$~km~s$^{-1}$ interior
to $\Delta r_{\perp} < 5 \times 10^3$~AU. This is to exclude the two extreme systems featuring HD~1917 and HD~214165, both of which have 
$\Delta V >8$~km~s$^{-1}$ but {\tt ruwe}$>1.4$. Ultimately this leaves us with an
 initial sample of 446 stellar pairs.

\begin{figure}
\includegraphics[height=6cm, width=5cm, angle=0, scale=1.75]{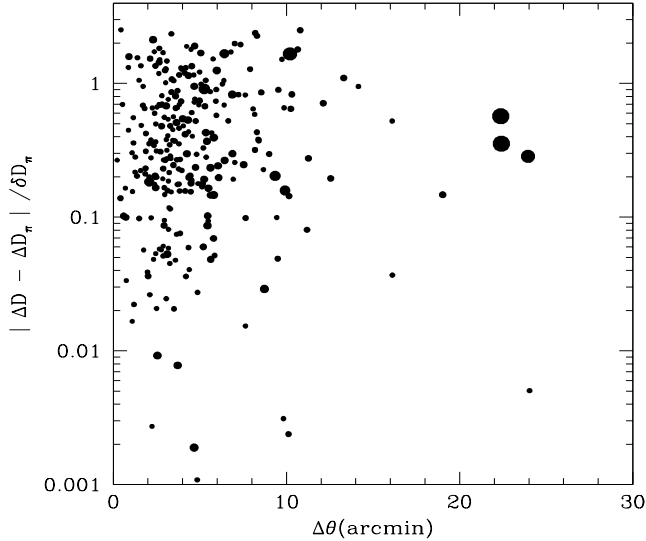}
\caption{   The quantity $\Delta D$ represents the difference in distance to two members of a close pair, as
calculated using the model of \cite{BJ21}. The quantity $\Delta D_{\pi}$ indicates the corresponding radial separation
using the simple conversion from parallax $D = 1000 pc/\pi$ and $\delta D_{\pi}$ is the distance
error based on the parallax measurement -- averaged over the two members of the binary. $\Delta \theta$ is
the angular separation on the sky for each pair. The size of each point scales as the square root of the parallax
to each source. This demonstrates that the change in radial separation is small in many cases, but does become
significant -- relative to the nominal error -- in some cases. \label{Ddif2}}
\end{figure}

\begin{figure}
\includegraphics[height=6cm, width=5cm, angle=0, scale=1.75]{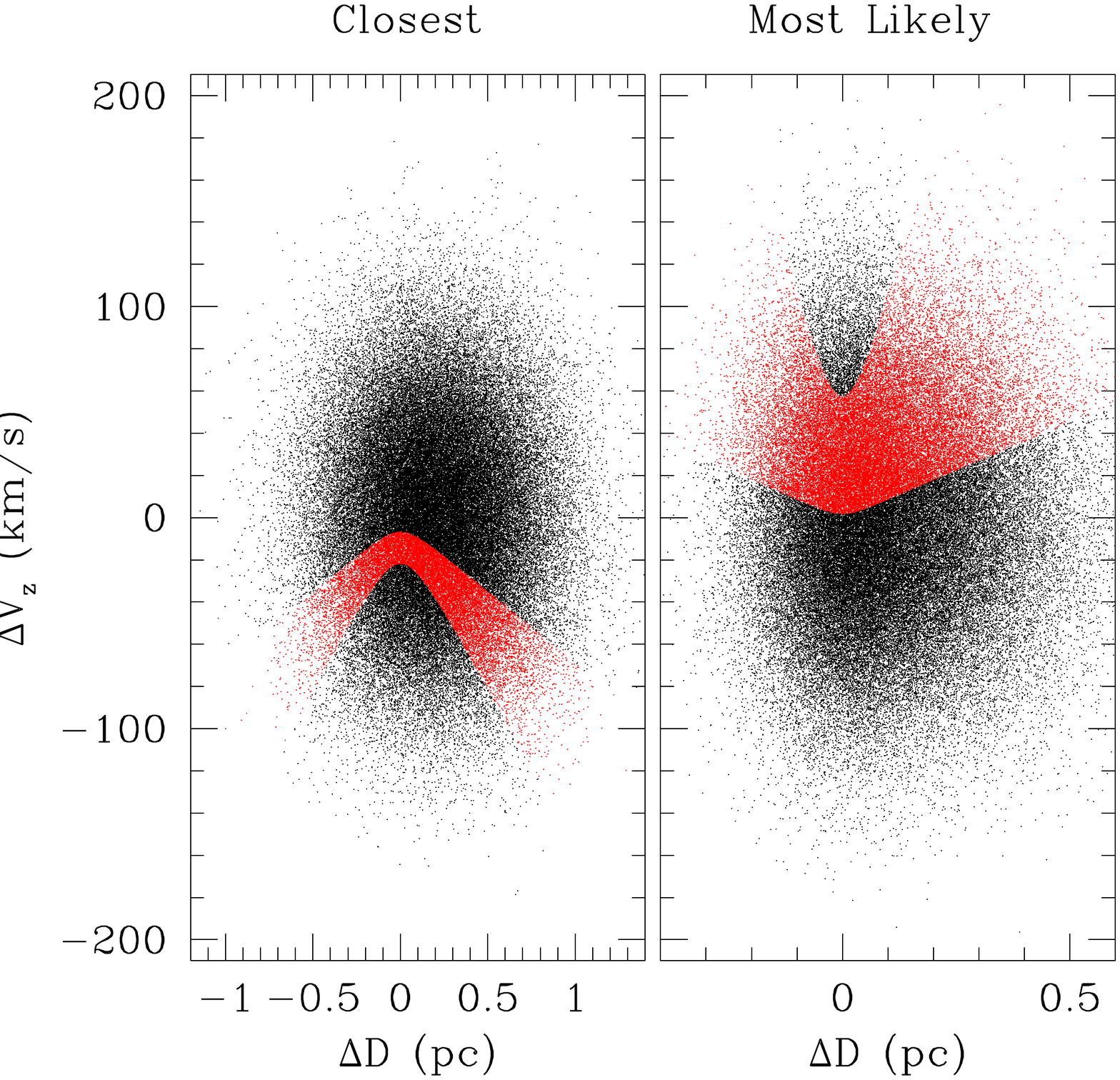}
\caption{The black points show the choices of relative radial distance and relative
radial velocities, for a Monte Carlo simulation of the possible trajectories for two pairs in
our sample.
The distances  are chosen using
the measured uncertainties from \cite{BJ21}, and the relative radial velocities  are drawn from
the distributions in the Galactic model of \cite{Ryb20}.
The red points indicate those trajectories that allow the two stars to pass within $10^4$~AU of
each other. On the left, we show the results for the pair in our sample that allows for the closest possible
passage, namely the pair  2MASS~J07315294+0633251 and 2MASS~J07321651+0628319.
On the right, we show the results
for the pair that has the greatest probability of having a passage within $10^4$AU, namely 
2MASS~J05100438+3306067  and 2MASS~J05100212+3303535.
 In neither of these two cases do we know the radial velocities
of the stars in question.
\label{ddvz}}
\end{figure}

The third, radial, component of the separation is the one with the largest error,  even with the unprecedented
accuracy of the GAIA parallaxes. Furthermore, the systematic errors in the astrometry can vary depending on how
crowded the field is. The angular separations of the pairs considered here can extend to $>20'$, and so we adopt the
corrected distances from \cite{BJ21} to ensure a uniform treatment for our entire sample.
 \cite{BJ21}
provide improved distance estimates to GAIA sources using two methods. The geometric method adopts a prior distribution
based on the Galactic distribution of stars in that direction, while the photogeometric estimate includes
information from the photometry of the stars in question, and their expected position in the color-magnitude
diagram.  We therefore calculate the probability distribution of true separations for our stellar pairs using
the photogeometric distance distributions from Bailer-Jones et al.  Figure~\ref{Ddif2} shows the difference
between the revised distances and the naive, parallax-based distance $D_{\pi}=1000 AU/\pi$, where $\pi$ is the
parallax in units of milliarcseconds. While the difference is well within the errors for many systems, there is also
a notable fraction for which the correction is important.

In order to determine the true distance of closest approach in each encounter, we need to specify
the one remaining component -- the relative radial velocity. While GAIA does provide radial velocities for
a subset of stars, most of the ones in question here are not part of that sample. To determine the
radial velocities, we sample the radial velocities contained in the Galactic model used by \cite{BJ21} to
determine the distance probabilities. This model is described in \cite{Ryb20}, and we query the resulting mock catalog
for each pair. For each object,
we draw all radial velocities for stars with $2^{\circ}$ of the position to determine the radial velocity distribution. We also restrict our query to model stars within $\pm 5$ magnitudes of the
object. If two members of a pair have significantly different magnitudes, they may then have slightly different line-of-sight velocity dispersions, but there
are no cases with large discrepancies.
%
For each choice of radial velocity, we then project a straight-line trajectory forwards and backwards in time and
isolate the distance and time of closest approach. The encounters we consider here are significantly hyperbolic and so
deviations from straight trajectories are negligible. For each pair, we evaluate $10^7$ trajectories, sampling the
relative distance and relative velocity distributions determined above. Our interest is in those systems which pass the closest,
as these will be the ones that would facilitate interstellar migration. To score the relative attractiveness of each system, we use two metrics.
In the first, we  count the number of trajectories that result in a passage
closer than $10^4$~AU  and quantify the fraction  $P_{\mu}$ of all resulting trajectories that pass within this distance.  A second metric is to
use the average $\langle 1/r_{min}^2 \rangle $, where $r_{min}$ is the minimum approach distance for each trajectory, and the average is performed
over all trial trajectories. The rationale for this metric
is that we want to select those systems that require the least energy expenditure per unit mass for a transfer between the stars. A smaller $r_{min}$
will require a smaller velocity to cross it in a finite time, so that we anticipate the energy will scale $\propto r_{min}^2$. To favor the
minimum energy, we therefore seek to maximize the average of $1/r_{min}^2$ over all our trajectories.

 For inclusion in our sample, we require $P_{\mu}>0.01$ or $\langle 1/r_{min}^2 \rangle >0.1$ (where $r_{min}$ is normalised to $10^4$AU). We have
115 pairs which satisfy both criteria, 13 which satisfy the first only and 9 which satisfy the latter only. Taking these together, our final sample
contains 137 of the original 446 pairs.
We also find that  30 of these pairs involve at least one star that has a possible bound companion in the GAIA
sample. In some cases we find that both members of the binary are counted independently as having an encounter with the
third star. After accounting for these duplications, we find  132 independent encounters,  28 of which involve multiple stars.
Of these,  four encounters feature pairs in which both members have proper motion companions.

 We have also examined the {\tt ruwe} parameter for the 269 individual stars involved in the systems of our final sample, and find that
38 of them exhibit a value $>1.4$, suggesting that about 14\% of our stellar sample contains partially resolved binary stars.
This is a representative value for the stellar population as a whole, and reinforces our conclusion -- in Figure~\ref{Ddif2} -- that
the upper envelope of the bound population is composed primarily of hierarchical triples, because of the much higher fraction with {\tt ruwe}$>1.4$.

Figure~\ref{ddvz} illustrates  the process by which we estimate the metrics $P_{\mu}$ and $\langle 1/r_{min}^2\rangle$. 
We show two examples. On the left, we show
 the case of the closest passage found in our sample -- between the pair of stars  
 2MASS~J07315294+0633251 and 2MASS~J07321651+0628319.
 The proper motions and astrometric separations are also sampled with their assumed errors, although their uncertainties have
 a small effect compared to the radial distance uncertainty. The right panel shows the case where we find the largest probability of
 an encounter within $10^4$AU, namely the pair  2MASS~J05100438+3306067  and 2MASS~J05100212+3303535.
 The reason for the
 higher probability can be seen from the dimensions of the x-axis -- the estimated distances for these two stars are much closer to one another than most
 of the other stars in the sample (their mean distances are only  0.07~pc discrepant).
 

\subsection {Encounters involving Sun-like Stars}

We are particularly interested in close passages featuring Sun-like stars, as the motivation for this sample is
to identify cases where Earth-analogue technological civilizations might attempt to migrate to another star. Therefore,
Table~\ref{TabG} collects the information for that subset of the full sample of  132 encounters which feature  at least one star with $M_G<6$,
where $M_G$ is the absolute magnitude in the broad GAIA passband G. The spectral
classifications (taken from the literature) for the resulting stars range from K1--F3, and so these represent an approximately Sun-like sample.
The color $B-R$ represents the $G_{BP}-G_{RP}$ color obtained by integrating the GAIA low resolution blue and red prism spectra. Distances are the photogeometric distances from Bailer-Jones et al. (2021).  
The time of closest approach is relative to today -- positive values mean that the closest approach will occur in
the future, while negative values indicate it already occurred. Co-ordinates for the brighter member of each 
 pair are given in the finding charts in the Appendix.

\begin{splitdeluxetable*}{lllccclllllBlllllll}
\tablecaption{ \label{TabG} The sample of close, unbound pairs featuring Sun-like stars, ordered in increasing minimum possible close approach distances.
}
\tablehead{\colhead{Index} & 
\colhead{$\Delta r$} & \colhead{$T_{min}$} & \colhead{$P_{\mu}$} & \colhead{$P_{\rm rv}$}& \colhead{$\langle \left(\frac{10^4 AU}{\Delta r}\right)^2 \rangle $} & \colhead{Name (Type)} & \colhead{ $D$ }& \colhead{$ M_G$} & \colhead{ $B-R$} & \colhead{ $V_{\rm rad}$} & \colhead{Index} & \colhead{Name (Type)} & \colhead{ $D$ }& \colhead{$ M_G$} & \colhead{ $B-R$} & \colhead{ $V_{\rm rad}$} & \colhead{Notes} \\ &
\colhead{(AU)} & \colhead{(yrs)} & &  & &  & \colhead{(pc)} & & & \colhead{(km\, s$^{-1}$)} &  & & \colhead{(pc)} & &  & \colhead{(km\, s$^{-1}$)} & 
}
\startdata
\hline 
1 & 1367 & +1750 & 0.05 & 0.03 & 0.55 &  HD 157393  (F8) & 59.43$^{+0.17}_{-0.20}$ & 3.74(2) & 0.65 & -21.9(4) & 1 & LSPM J1722+1513 & 59.15$^{+0.09}_{-0.09}$ & 10.73(1) & 2.70 &  &  (1,4) \\
\hline 
2 & 1413 & -153 & 0.02 & 0.02 & 0.20 & HD~159902 (G5V) & 49.55$^{+0.22}_{-0.27}$ & 4.81(2) & 0.82 & +5.4(4)  & 2 & CD-45 11747 & 49.11$^{+0.03}_{-0.04}$ & 5.95(1) & 1.07 & -32.0(5) & (1,2,4) \\
\hline 
3 & 2032 & -4194 &  0.14 & 0.02 & 1.07 & HD 187154 (G1V) & 44.98$^{+0.05}_{-0.06}$ & 4.38(1) & 0.78 & -32.3(2) & 3 & HD 187084 (K1V) & 45.27$^{+0.05}_{-0.05}$ & 5.64(1) & 1.01 & +9.5(2) & (1) \\
\hline 
4 & 2957 & + 15250 & 0.15 & 0.20 & 0.63 & HD~87978 (G6IV) & 40.84$^{+0.06}_{-0.05}$ & 4.92(1) & 0.84 & 28.7(2) & 4 & J10090601-1109290 & 40.60$^{+0.6}_{-0.4}$ & 13.5(1) & 4.12 & & (5)\\
\hline 
5 & 3193 & +1600 & 0.01 & 0.03 & 0.08 & HD~92577 (F2V) & $94.66^{+0.32}_{-0.25}$ & 3.14(2) & 0.51 & -8.9(6) & 5& J10411261-1418509 & $93.99^{+0.86}_{-0.73}$ & 12.09(4) & 3.06 &  & \\
\hline 
6 & 3849 & +3000 & 0.12 & $<0.01$ & 0.55 & TYC~5786-123-1   & $87.93^{+0.11}_{-0.11}$&  5.89(1) & 1.14 &  +29.0(7) & 6 & J21335912-0746358 & $87.86^{+0.64}_{-0.52}$ & 12.59(3)  & 3.36 & & (4,6) \\
\hline 
7 &  6341 & +10152 & 0.10 & 0.14 & 0.28 & HD 143332 (F5V) & 99.07$^{+0.36}_{-0.40}$ & 2.88(2) & 0.70 & -29.8(2) & 7 & J16000455-0712125 & 99.95$^{+0.79}_{-0.70}$ & 11.23(4) & 2.97 &  & (1) \\
\hline 
8 & 6374 & +3994 & 0.10 & 0.09 & 0.28  & HD 26770 (G0V) & 64.11$^{+0.12}_{-0.14}$ & 3.99(1) & 0.70 & +17.3(6) & 8 & J04132532-2830440 & 63.90$^{+0.15}_{-0.13}$ & 11.96(1) & 3.12 & & (2) \\
\hline 
9 & 6402 & -151 & 0.06 & 0.07 & 0.19 & HD 50669 (G8IV) & 94.17$^{+0.13}_{-0.13}$ & 3.87(1) & 0.86 & +73.0(4) & 9 & J06540919-0148118 & 93.86$^{+0.37}_{-0.43}$ & 10.94(2) & 2.48 &  & \\
\hline 
10 & 6739 & +2538 & 0.07 &  & 0.20   & HD 44006  (G0V) & $96.04^{+0.18}_{-0.17}$ & 3.93(1) & 0.69 &  & 10 & J06191745-0020445 & $95.91^{+0.71}_{-0.51}$ & 8.71(3) & 2.26  &  & (5) \\
\hline 
11 & 7819 & -357 & 0.06 & 0.05 & 0.33 & HD 210974A/B (K0V) & 85.80$^{+0.17}_{-0.20}$ & 5.90(1) & 1.01 & -52.3(3) & 11 &  UCAC2 18317515 & 85.75$^{+0.13}_{-0.10}$ & 7.47(1) & 1.72 & -13.7(4) & (2) \\
\hline 
12 & 8774 & -6456 & 0.10 & 0.16 & 0.29 & HD 80790 (F6V) & 54.68$^{+0.05}_{-0.06}$ & 3.25(1) & 0.69 & -19.4(2) & 12 & J09191695-5901597 & 54.92$^{+0.22}_{-0.22}$ & 11.08(2) & 2.83 & & (7) \\
\hline 
13 &  8832 & +1195 & 0.07 & 0.05 & 0.37 & LSPM J2126+5338 & 53.48$^{+0.03}_{-0.03}$ & 5.97(1) &  1.10 &  -15.1(46) & 13 &  J21265700+5343001 & 53.50$^{+0.07}_{-0.10}$ & 11.28(1) & 3.01 &  & (3) \\
\hline 
14 & 9831 & -649 & 0.006 & $<0.01$ & 0.14 & LSPM J0646+1829 & 86.05$^{+0.10}_{-0.12}$ & 5.83(1) & 1.03 &  +75.2(8) & 14 & J06464503+1828391 & 86.13$^{+0.41}_{-0.36}$ & 11.37(2) & 2.88 &  & \\
\hline 
15 & 13033 & +1250 & 0 & 0 & 0.37 & HD 29836 (GV) & 42.22$^{+0.03}_{-0.04}$ & 3.84(1) & 0.83 &  13,3(1) & 15 &  LP 415-358 & 42.04$^{+0.09}_{-0.07}$ & 12.32(1) & 3.10 & & (2) \\
\hline 
16 & 26072 & +850 & 0 & 0 & 0.13 & HD 192486 (F2V) & 43.94$^{+0.04}_{-0.05} $ & 3.24(1) & 0.53 & -8.9(2) & 16 & HD 192527(G5V) & 44.02$^{+0.05}_{-0.04} $& 5.07(1) & 0.86 & -8.8(2) \\
\enddata
\tablecomments{ 
(1) Proper motion anomaly between HIPPARCOS and GAIA DR2 suggests binarity for the brighter member of this pair;
(2) The brighter of the two members of this encounter has a proper motion companion;
(3) The fainter of the two members of this encounter has a proper motion companion;
(4) The brighter of the two members of this encounter shows excess astrometric noise
(5) The fainter of the two members of this encounter shows excess astrometric noise
(6) Both members of this encounter have proper motion companions;
(7) Identified as a binary in the literature;
}

\end{splitdeluxetable*}

There are also radial velocities available for many of the brighter stars in this sample, and these are listed in Table~\ref{TabG}. Of the  16 pairs in Table~\ref{TabG},
 15 have at
least one radial velocity available, and  four pairs have known radial velocities for both members. 
The radial velocities are drawn from the GAIA EDR3
data, except for those for HD~92577 and HD~26770, which are drawn from \cite{Pulkovo}.
In no case do we mix velocities from different references, and so we do not have to consider the possibilities of differences in
offsets or definition. 
 Using this additional information we have calculated a second probability
of encounter for these objects (P$_{\rm rv}$, as opposed to the  probability P$_{\mu}$, which is based purely on the measured proper motion). In the cases where both radial velocities are known, we do not need to assume any
distribution for line-of-sight velocities, but simply repeat our integration of trajectories using the known relative velocity and sampling only within the
quoted errors. In cases where the radial velocity is known for only one member, we sample the distribution of radial velocities for the other star,  
using the mock catalog of \cite{Ryb20} as described earlier. In general terms, the inclusion of this extra information lowers the probabilities of close encounter. In fact, after including
the radial velocities, 2/16 of our systems
drop below the 1~per cent probability threshold. The product  $\Pi_{i=1}^{16} (1-P_{\rm rv})_i = 0.37$ indicates that there is a  37~per cent chance that none
of these  16 systems have a close passage within $10^4$AU, or that there is a 63~per~cent chance that at least one system does. More detailed
examination of the combinatorics indicates that there is a  39~per cent chance of one system passing close enough, an 18~per cent chance that two pass close enough,
and a  6~per cent chance that more than 2 of these systems will pass within $10^4$~AU of each other.

\begin{figure}
\includegraphics[height=6cm, width=5cm, angle=0, scale=1.75]{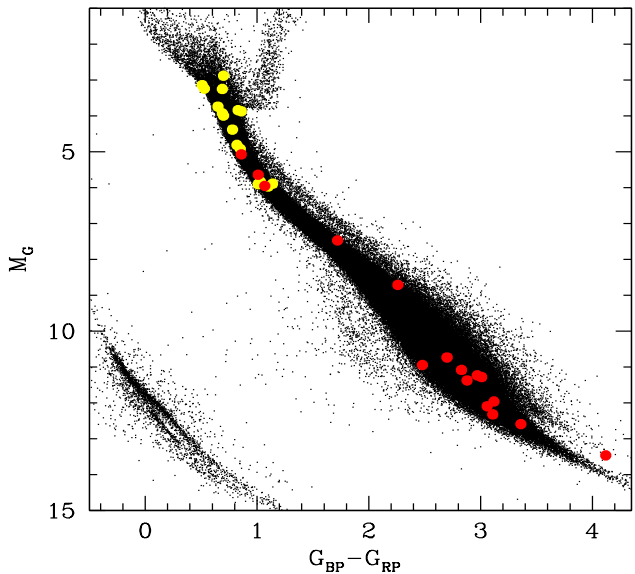}
\caption{ The small black points represent the population of well measured stars from GAIA within
100~pc of the Sun. In this context, well measured means fractional parallax errors of less than 10\%,
5$\sigma$~astrometric error parameter $<$0.1 and photometric precision in G $<0.02$. The yellow points represent the brighter members of
pairs in which one of the stars has absolute GAIA magnitude $M_G<6$. The red points represent the fainter members of these pairs.
 \label{CMD}}
\end{figure}

Figure~\ref{CMD} shows the photometry for these  16 pairs in Table~\ref{TabG}.
 The photometry for the brighter (yellow) and fainter (red) members of
these pairs are compared to that of the field sample within 100~pc. As expected, the encounters are
primarily with lower mass stars, although there are a few pairs featuring similar stars.

Another important element of the HZ21 argument is that close stellar encounters represent the
best chance for migration in the case of technological civilizations that orbit single stars. If the planet orbits a star
in a multiple system, then the bound companions are much easier targets and more likely to be long-lived, given
the stellar demographics. To that end, it is important to constrain the binarity of the bright stars in our sample.

Indeed,  four of the systems in Table~\ref{TabG} feature  a bright star which has a high probability proper
motion companion in the GAIA EDR3 database.
In cases where our search independently flagged encounters with different members of binaries, we report only
the encounter featuring the brightest star (so that each overall encounter between bound systems is counted only once).
 Furthermore,  four stars also show a proper motion anomaly between the measurements of HIPPARCOS and
GAIA DR2 \citep{Kervella}, which is taken to be a signature of binarity as well. Finally, one more system is identified
as a binary in the literature \citep{Tokov}. Six of the stars in this sample also have {\tt ruwe} $>1.4$, which may also
indicate binarity.
 Overall,  8/16 of the bright stars
in this sample show evidence for binarity. 
Furthermore, this census is likely to undercount companions, as our constraints are primarily for wide binary members
and closer companions may still remain undetected.

\subsection{The lower mass sample}

We have compiled the rest of the likely close encounters (those that feature no stars with $ M_G<6$) in Table~\ref{TabM}.
In this table we calculated only the probability of encounter on the basis of proper motions (P$_{\mu}$) as radial velocities
are not available for most of the stars in the lower mass sample. This leaves us with  116 encounters, to go with the  16 in the
bright star sample.

\begin{figure}
\includegraphics[height=6cm, width=5cm, angle=0, scale=1.75]{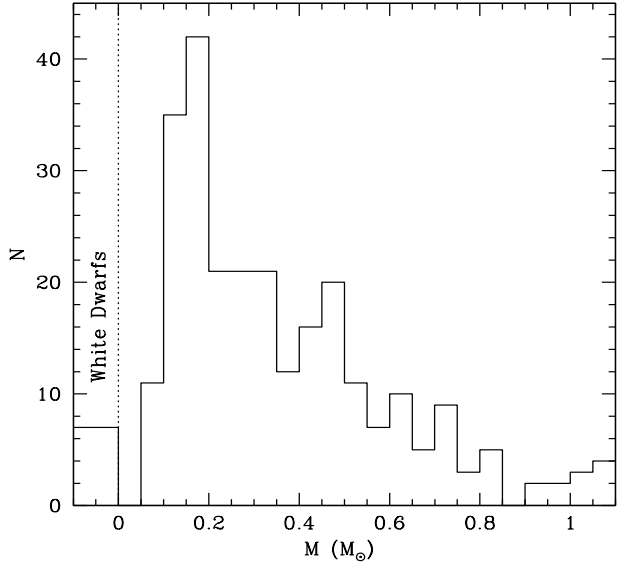}
\caption{ The histogram shows masses estimated for the stars in Tables~\ref{TabG} and \ref{TabM}, based on
the GAIA magnitudes and colours, and using stellar models and colours from \cite{MIST16}. The bin at
negative stellar mass represent the white dwarfs in the sample. White dwarfs of all masses are counted in this bin.
 We see that the sample is dominated by the lowest mass stars.
 \label{Mhist}}
\end{figure}

With this sample we can also examine the overall demographics of the stars involved in these close encounters. The
mass distribution of the stellar population from Tables~\ref{TabG} and \ref{TabM} is shown in Figure~\ref{Mhist}.
 Stellar masses are based on the
position of each star in the GAIA magnitude system, and a comparison with the MESA models \citep{MIST16}.  Seven of
the pairs contain a white dwarf member. The result, as expected,  is strongly peaked between $0.1$--0.2$M_{\odot}$, and appears quite consistent with a random sampling of the field stellar mass function.

 With the larger number of systems, here,  the probability that none will  pass with $10^4$AU is negligibly small ($\sim 7 \times 10^{-7}$).
A perhaps more practical logistics question is to explore how many of these systems need one explore to have a realistic chance of seeing
a passage within $10^4$~AU. If we calculate probabilities in Table~\ref{TabM} starting from the first entry, we need only go to the  twentieth entry to 
have a 50\% chance of two or more close passages. This is consistent with the order of magnitude estimate that 
 about 1 in 10
of the systems in Tables~\ref{TabG} and \ref{TabM} are expected to actually pass within $10^4$~AU of each other.
 How does this compare
to our expectations? If we adopt the local stellar encounter rate calculation in the Solar neighbourhood as
estimated in HZ21, using a local stellar density $n_*$ and assuming the dispersion in relative velocities is
$\sqrt{2}\sigma$, where $\sigma$ is the local average velocity dispersion in the Galactic disk, 
we estimate the rate of encounter, per star, is
\begin{equation}
\Gamma = 0.016 \, {\rm Myr^{-1}} \left( \frac{n_*}{0.1\, {\rm pc}^{-3}} \right) \left( \frac{R_0}{10^4 {\rm AU}} \right)^2 \left( \frac{\sigma}{48 \, \rm km s^{-1}} \right)
\end{equation}
If we estimate time of encounter to be $\sim 2 R_0/V \sim 1400$~years, then we estimate $\sim 9$ ongoing encounters within 
the volume of 100~pc of the Sun. This suggests our sample contains a significant fraction of the close encounters happening
in this volume.

\begin{figure}
\includegraphics[height=6cm, width=5cm, angle=0, scale=1.75]{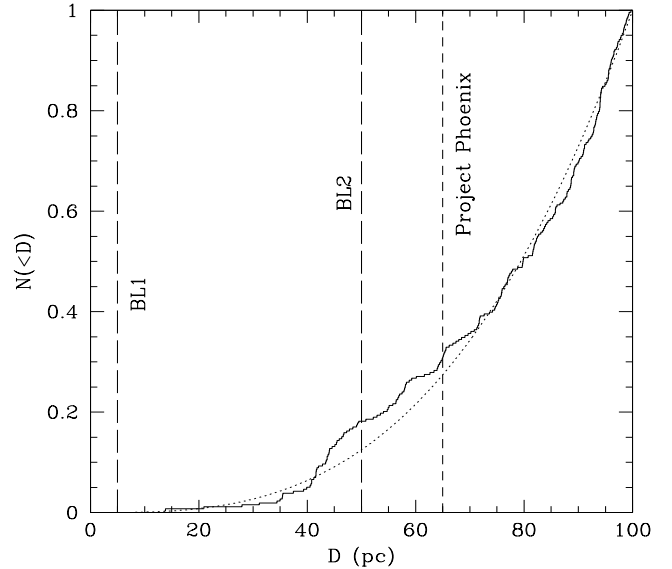}
\caption{ The histogram shows the cumulative distribution of counts of stars versus distance in our
sample. The dotted line indicates the function $P(D) = (D/100 \, {\rm pc})^3$, which represents the expected distribution
for a volume complete sample of uniform density, out to 100~pc.  The short dashed line indicates the distance limit
of the Project Phoenix search for Technosignatures, while the two long dashed lines indicate the two sub-samples of the
original Breakthrough Listen project \citep{Isaac17}. \label{Dist}}
\end{figure}

 Figure~\ref{Dist} shows the distribution of distances from the Sun shown by the cumulative  sample in Tables~\ref{TabG} and \ref{TabM}.  We see
that the bulk of our targets lie at distances farther than the samples \citep{Isaac17} used by the stellar surveys of Project Phoenix (65~pc) and Breakthrough Listen (5~pc --complete, and 50~pc --representative).
The distribution of the sample with distance  scales approximately with the enclosed volume out to our distance limit of 100~pc. This is consistent with
estimates 
 of GAIA EDR3 completeness \citep{Fab21,Brown21,Smartt21} , which suggest that the survey is highly complete between the magnitude limits $20<G<3$, albeit with
some degredation in crowded regions of the sky. Our use of a Galactic model to refine our distances helps
to mitigate the dangers of crowded regions and none of the stars in our sample are bright enough to saturate the GAIA photometry.  Furthermore, the separations of interest to us
are much larger than the angular scales at which GAIA blends the images of neighbouring stars -- see Figure~\ref{Ddif2}.

\begin{longrotatetable}
\begin{deluxetable*}{lccclcccccccl}
\tablehead{
\colhead{Name (Type)} & \colhead{ $D$ }& \colhead{M$_{\rm G}$} & \colhead{ B-R} & \colhead{Name} & \colhead{$D$} & \colhead{M$_{\rm G}$} & \colhead{B-R} & 
\colhead{ $\Delta r$} & \colhead{ $T_{min}$ }& \colhead{Prob} &  \colhead{$\langle \left( \frac{10^4 AU}{\Delta r} \right)^2\rangle$}  &\colhead{Notes}   \\
\colhead{2MASS} & \colhead{(pc)} &     &  & \colhead{2MASS} & \colhead{(pc)} &   &    & \colhead{ (AU)}  & \colhead{(yrs)} &  & & \\
}
\tablecaption{\label{TabM} The sample of close, unbound pairs featuring lower mass stars, ordered in increasing minimum possible close approach distances.
In cases where the star is not included in the 2MASS database, identifiers are either from SDSS (start with "S") or from the Gaia
DR2 catalog (start with "G"), followed by an index. The names themselves, in these cases, are too long to fit the column width and so are given in
the notes at the end.
Spectral
types are indicated in the notes with the notation [1/2], when known, and those members of the encounter pair that are members of multiple systems are indicated
by (bin-i) in the notes section, where i indicates either the first or second entry in the row.  
}
\startdata
\hline 
J07315294+0633251 & 56.46$^{+0.05}_{-0.05} $ & 9.58(1) & 2.46 &
J07321651+0628319 & 56.28$^{+0.44}_{-0.47} $ & 11.74(4) & 3.29 & 22 & +10050 & 0.17 & 13.2 &  \\
\hline 
J22314220+0023112 & 96.06$^{+0.34}_{-0.30}$ & 10.24(2) & 2.65 &
J22313478+0027179  & 96.55$^{+1.35}_{-1.73}$ & 12.97(9) & 3.18 & 118 & -3701 & 0.03 & 2.73 & [$\cdots$/M4V] \\
\hline 
J01024057+1610560 &    93.38$^{+0.37}_{-0.37} $ & 10.03(2) & 2.43 &
J01023204+1612129 &   92.92$^{+0.51}_{-0.53}$ & 10.76(3) & 2.72 & 171 & -4601 & 0.10 & 6.09 &  \\
\hline 
G1 & 55.29$^{+0.58}_{-0.46}$ & 14.87(5) & 1.01  & 
TYC 1568-917-1 & 54.88$^{+0.05}_{-0.05}$ & 7.62(1) & 1.65 & 381 & +2593 & 0.09 & 3.35 &   \\
\hline 
S1 & 45.54$^{+0.28}_{-0.28}$ & 14.84(3) & 0.93 & 
J00352199+2006142 & 45.45$^{+0.09}_{-0.07}$ & 11.88(1) & 3.04 & 398 & -2698 & 0.14 & 5.20 &  (bin-1)\\
\hline 
J19200086-2241531 &   94.27$^{+0.46}_{-0.46}$ & 10.68(2) & -0.24 &    
J19195587-2242109  &  92.92$^{+0.50}_{-0.53}$ & 11.42(3) & 2.91 & 451 &  +950 & 0.09 & 3.11 & (bin-1) \\
\hline 
J07084492+5544576 &   97.31$^{+0.63}_{-0.77}$ & 10.26(4) & 2.48 &
J07085805+5545261 &   96.95$^{+1.33}_{-1.37} $& 12.85(7) & 3.41 & 498 & -2950 & 0.04 & 1.22 &   \\
\hline 
GD 84  & 35.45$^{+0.04}_{-0.04}$ & 12.41(1) & 0.103 & 
LP 162-1  & 35.33$^{+0.02}_{-0.03}$ & 10.62(1) & 2.85 & 522 & +4034 & 0.20 & 5.76 & (1),[WD/M4] \\
\hline 
J19545686-6724041 & 57.15$^{+0.43}_{-0.39}$ & 9.06(4) & 2.29 &
J19544917-6719559 & 56.91$^{+0.28}_{-0.24}$ & 8.82(3) & 2.12 & 587 & +1148 & 0.04 & 0.95 & (2),(bin-1) \\
\hline 
J06433851-6417204 &  $93.19^{+1.38}_{-1.19}$ & 10.76(6) & 2.97 &
J06434304-6417291 &  $92.84^{+0.18}_{-0.13}$ & 9.85(1) & 2.47 & 637 & -845 & 0.04 & 1.00 & \\
\hline 
J06363407+0806450 & 40.85$^{+0.09}_{-0.08}$ & 13.08(1) & 3.50 &
J06364675+0808373 & 40.60$^{+0.04}_{-0.05}$ & 11.48(1) & 2.58 & 662 & -1897 & 0.12 & 2.72 \\
\hline 
J10520964-0548506 &  91.19$^{+0.72}_{-0.62}$ & 11.91(4) & 3.11 &
J10515946-0549049 &  90.87$^{+0.48}_{-0.45}$ & 10.65(3) & 2.72 & 738 & +898 & 0.02 & 0.17 & \\
\hline 
J09051507-2342573  & $88.57^{+0.32}_{-0.29}$ & 11.85(2) & 3.13 &
J09051785-2341229 & $88.68^{+1.24}_{-1.09}$ &  8.41(6) & 1.97 & 810 & +744 & 0.03 & 0.49 \\
\hline 
J08245980-1745049 & 89.29$^{+0.56}_{-0.59}$ & 12.18(3) & 2.52 &
J08244862-1741209 & 88.65$^{+0.73}_{-0.73}$ & 12.50(4) & 3.15 & 976 & -3892 & 0.04 & 0.60 & (bin-1) \\
\hline 
J05295170-3006557 &  98.46$^{+0.62}_{-0.500}$ & 12.07(3) & 3.24 &
J05295196-3003595 &  98.31$^{+0.58}_{-0.62}$ & 12.39(3) & 3.35 & 1048 & -2951 & 0.07 & 1.10 & \\
\hline 
J04590411+3804240 & 77.42$^{+0.94}_{-0.92}$ & 13.40(6) & 3.58 & 
J04592452+3801272 &  76.95$^{+0.28}_{-0.21}$ & 11.17(2) & 2.65 & 1074 & +2902 & 0.05 & 0.66 & (bin-2)\\
\hline 
J08483006-0745120 &  81.70$^{+0.35}_{-0.33}$ & 11.64(2)& 3.10 &
J08482010-0747371 &  81.65$^{+0.16}_{-0.14}$ & 7.51(1) &  1.67 & 1110 & -5652 & 0.20 & 2.23 & \\
\hline 
J05381466+3056067 & 75.01$^{+0.52}_{-0.56}$ & 12.39(4) & 3.50  & 
J05380946+3055453 & 74.63$^{+0.11}_{-0.10}$ & 8.65(1) & 2.02 & 1196 & -1298 & 0.12 & 1.57 \\
\hline 
J06300702+7343471 & 98.73$^{+2.0}_{-1.65}$ & 13.9(1) & 3.60 & 
J06302012+7342514 & 97.79$^{+0.47}_{-0.54}$ & 9.52(3) & 2.32 & 1215 & -4745 & 0.04 & 0.51 \\
\hline 
TYC~5133-940-1 & 94.13$^{+0.14}_{-0.14}$ & 6.27(1) & 1.22 & 
 J19134957-0217557  &  94.04$^{+0.59}_{-0.51}$ & 11.89(3) & 3.24 & 1272 & +4398 & 0.20 & 0.38 & (bin-1) \\
\enddata
\tablecomments{ 
 (1) GD~84 is referenced in papers about low mass companions 
(2) These sources are in a very confused region in DSS/2MASS. 
G1=Gaia DR2 4522020089271002496; S1=SDSS J003542.03+200938.4
{\em This table shows the first twenty entries.
The full table is available in machine readable format online.}
}
\end{deluxetable*}
\end{longrotatetable}

\subsection{Notes on Individual Systems}

\begin{figure}
\includegraphics[height=6cm, width=5cm, angle=0, scale=1.75]{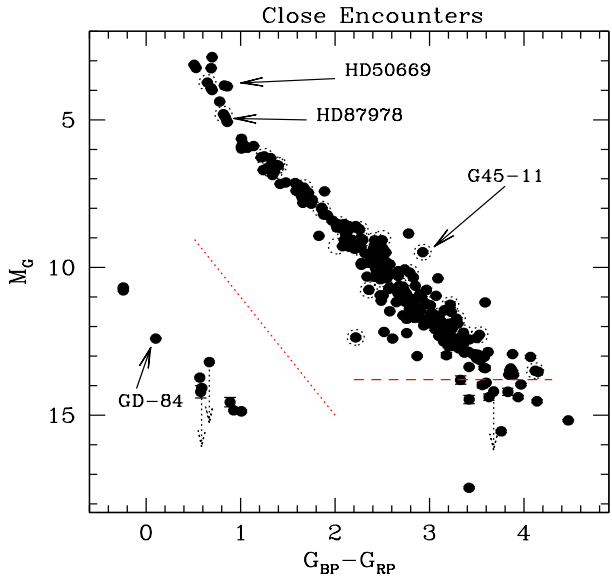}
\caption{ This shows the stellar parameters of all the stars involved in close, unbound, stellar encounters, from Tables~\ref{TabG} or \ref{TabM}.
The red horizontal dashed line indicates an approximate dividing line between true hydrogen burning stars and brown dwarfs (although the exact location would
be age and metallicity dependant). The diagonal dotted line indicates the division between white dwarfs and main sequence stars. A handful of noteworthy objects
are indicated, and discussed in the text.
 \label{NewSample}}
\end{figure}

The  132 encounters identified here involve a variety of stellar systems. Figure~\ref{NewSample} shows the GAIA bandpass colour-magnitude
diagram for the complete set. A few notable systems are

\begin{itemize}
\item  HD87978 is classified as G6IV and so should be starting to leave the main sequence, and therefore is the perfect archetype of the rationale in HZ21 for a system
in which a civilisation might be motivated to make
an attempt at interstellar migration. It is also the system in Table~\ref{TabG} with the highest probability of an encounter within $10^4$ AU. One potentially
complicating factor is that the star with which it is having an encounter appears to show evidence for astrometric noise and so may be a partially resolved
binary. How this influences its suitability as a migration destination will depend on the actual parameters of the system. \\
\item The star HD50669 is classified as G8IV and does indeed lie slightly to the red of the upper main sequence. This star is  also
starting to leave the main sequence.  This system also shows no sign of binarity. Unfortunately, given the uncertainties in distances, it has only
a 3~per cent chance of actually passing within $10^4$~AU. \\
\item A comparison of the sample in Table~\ref{TabG} with the NASA Exoplanet Archive \citep{Nexsci} shows no known exoplanets around these stars.
One star -- HD143332 -- is included  in the Lick-Carnegie Radial velocity sample \citep{Butler17}, but has no reported planet. It does show a non-zero
acceleration term, however. None of the sample is recorded in the archives of the  HARPS survey \citep{HARPS20} or the California Legacy Survey \citep{CLS21}. This paucity is not entirely surprising, as the radial
velocity searches are strongly weighted towards nearby stars, and most of the stars in this sample are too distant to be well represented. \\
\item On the lower main sequence, the star G45-11 lies above the main sequence, but is also a known spectroscopic binary and certainly lies in
the correct location for a binary unresolved by the GAIA photometry. As the lower mass member of an encounter featuring a K4V star (HIP 53175)
this may still be a viable candidate for a migration event, if  a planetary system formed around the higher mass star. Four stars lie above the lower
main sequence in Figure~\ref{NewSample} -- three are suspected binaries. The fourth (2MASS J13344645-5959405) lies in a region of the sky where source confusion is a significant
concern. \\
\item The stars to the left of the diagonal dotted line in Figure~\ref{NewSample} are likely to be white dwarfs. These are less likely to be viable
candidates for a migration because they do not offer a long-term stable habitable environment. However, none of the white dwarfs identified here
are in encounters with G stars in Table~\ref{TabG}. They are all undergoing encounters with M stars, so could potentially be the starting points for
migration from civilisations that have survived the ravages of stellar evolution. \\
\item One white dwarf system of note is that of GD-84, identified as a candidate wide binary system by \cite{FBZ} but only tentatively, because the
proper motion measurement was compromised by confusion with a background star. The GAIA proper motion is now able to rule this out as a 
bound companion. However, the original candidate companion is not the other star in this candidate encounter. We show a finding chart for this
field in the Appendix. \\
\item The horizontal dashed line in Figure~\ref{NewSample} indicates the approximate location of the base of the Hydrogen burning sequence 
(this shows the $M_G$ for a solar metallicity $0.1 M_{\odot}$ star of age 4~Gyr). Objects fainter than this are candidate brown dwarfs and also
not the optimal targets for migration because of their slowly decreasing luminosities. None of the encounters with Sun-like stars (Table~\ref{TabG}) feature
brown dwarf candidates.  Some of these candidates (and a couple of white dwarf candidates) have uncertain $M_G$ because the photogeometric distance estimates 
of \cite{BJ21} have significant tails to large values for these objects. \\
\item The closest system to the Sun in this sample is the encounter between two M dwarfs, HD331161B and G125-30, which are at $\sim 14$~pc.
The star HD331161B is part of a binary. \\
\item The encounter with the closest minimum distance is between another pair of M dwarfs, 2MASS~J07315294+0633251 and 2MASS~J07321651+062839, which
has a minimum encounter distance of  22~AU.  This is also the system with the highest score in the metric $\langle 1/r_{min}^2 \rangle$. \\
\item The encounter with the highest probability of passage within $10^4$~AU is the pair  2MASS~J05100438+3306067  and 2MASS~J05100212+3303535, which
has a probability of  37~per cent. The high probability here is driven by the very close agreement in the distances to these stars.
\end{itemize}

\subsection{Planet Hosting Stars}

Our close encounter sample defined above does not contain any stars known to host planets. Indeed, it contains
very few stars that have even been searched for evidence of planets. Therefore, let us examine the sample
of stars that are known to host planets, and estimate how close of an encounter is likely to occur amongst this
subsample.

\begin{figure}
\includegraphics[height=6cm, width=5cm, angle=0, scale=1.75]{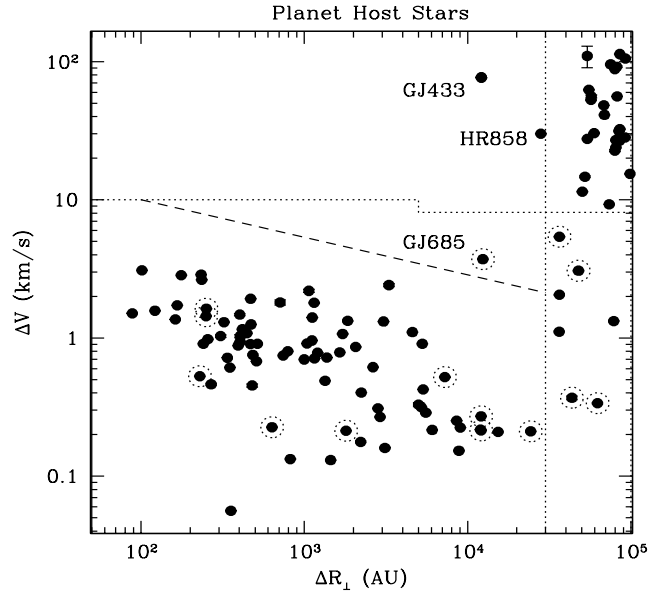}
\caption{ The solid points show matches between known planet-hosting stars and nearby
companions on the sky. The horizontal dotted line represents our nominal division between
bound (lower left) and unbound (upper right) pairs.  The region between this criterion and
the diagonal dashed line is suspected to contain mostly hierarchical triples whose relative
velocities are contaminated by partially resolved orbital motion in one or both components.
 The vertical dotted line indicates $\Delta R_{\perp}= 3 \times 10^4$~AU. 
The dotted circles indicate pairs which are not noted as binaries in the exoplanet database, but which appear to
exhibit the necessary common motion. All other points below the dashed line and to the left of the dotted line are known binary systems.
 \label{drdv}}
\end{figure}

To address this, we have repeated the GAIA search described above, but used, as the source 
population for the comparison, the stars within 100~pc of the Sun that are known to host planets. We draw this sample from
the NASA Exoplanet Database \citep{Nexsci}, as of July 9, 2021. Figure~\ref{drdv} shows the relative 
motions on the sky of all pairs whose parallaxes are consistent, and with projected separations $50 \rm AU < \Delta R_{\perp} < 10^5 \rm AU$. The
lower limit is chosen to exclude matches between the same star, since the NASA database uses GAIA~DR2
positions and we are using GAIA~EDR3 positions, which sometimes give small offsets. This leaves us with
129~pairs.

Figure~\ref{drdv} once again shows the division into bound and unbound pairs, as in Figure~\ref{Split}. There
are quite a few planetary hosts in binary systems, as is well known. The vertical dotted line indicates the
cut we used for further consideration in our original sample.
The  two unbound pairs from this sample that satisfy this criterion 
feature the stars GJ~433 and HR~858. These are the only two planet hosts that make it into our initial sample of  446 encounters.
\begin{itemize}
\item The star GJ~433 is an M2 dwarf ($\sim 0.5 M_{\odot}$) hosting three planets, whose periods are 7.37, 36.06 and 5094~days \citep{GJ433_1,GJ433_2,GJ433_3}.
About 2000~years ago, it may have come to within $\sim 9 \times 10^4$~AU of the K0V dwarf HD~100623, which is actually a binary containing
a white dwarf companion. Both of these stars have measured radial velocities as well, so this is a reasonably well determined measurement
(the biggest uncertainty being the respective radial distances).  This star does not make it into our sample because, without the radial velocity information,
 the chance of the close encounter
coming within $10^4$ AU is less than 1~per cent, and $\langle 1/r_{min}^2\rangle = 0.01$. However, when we include the known radial velocities,
$\langle 1/r_{min} ^2 \rangle$ increases to 0.16, although $P_{rv}$ is still $<0.01$.\\
\item The star HR~858/HD~17926 is a 1.2 $M_{\odot}$ F star at a distance of 32~pc, also with three planets -- periods 3.59, 5.97 and 11.23~days \citep{HR858}.  This star also
has a low mass binary companion. In $\sim 9000$~years, it will also come within $\sim 8 \times 10^4$~AU of another binary pair, dominated
by the star TYC~7012-885-1. Once again, this encounter is well characterised because both members have measured radial velocities. As in the case of GJ~433, this
pair also has a less than 1~per cent chance of coming within $10^4$~AU,  and  $\langle  1/r_{min}^2\rangle = 0.06$ in this case, even with radial velocities
included.
\end{itemize}

There is one other pair in Figure~\ref{drdv} that doesn't match well with the bound pairs -- that containing the planet hosting star
GJ~685 \citep{GJ685}.  This falls within the intermediate range we identified in \S~\ref{Close} as likely contaminated by hierarchical
triples. Indeed,
  this star is projected to undergo an encounter with the spectroscopic binary HD~160269. The fact that
both stars share the same radial velocity suggests that this is  truly a bound pair, and the larger-than-expected differential proper
motion may be affected by the orbital motion of the binary. There are several planet-hosting stars, designated with dotted circles in
Figure~\ref{drdv}, which are listed as single in the Exoplanet database but seem to have wide companions, based on their positions in
this diagram. These possible binaries are GJ~685, HD~238090, HD~10647, HIP~38594, HD~39855, HD~8326, HD~221420, HD~23472, HD~24085, HD~31253 and WASP-29.
Also circled in Figure~\ref{drdv} are the systems HD~124330 and AU~Mic, which appear to lie between the elliptic and hyperbolic populations. In the case
of AU~Mic, this may be a consequence of its youth and membership in the $\beta$~Pictoris moving group.


\section{Signatures of Interaction}
\label{Sigs}

 We have identified a set of stars undergoing close, but unbound, stellar encounters.
 HZ21 noted that the energy burden of transfer between stars is considerably
eased during such encounters and  postulated that most interstellar migration will occur during such episodes. If such migrations are accompanied
by an increase in construction of spaceships and communication within a migrating population, then they might prove to be
fruitful targets for observations. To that end, we review here the possibility of infrared excesses in this sample. We also present, in the Appendix, finding
charts for the 16 systems in Table~\ref{TabG}.

\subsection{Infrared Excess}

One possible source of observable signals is the large-scale construction of spaceships to transport material and individuals from one star to the
next. The most energy efficient way of doing this would be to utilise a population of asteroids, and so we might search for
an infra-red excess resulting from dust released during the construction process.

\begin{figure}
\includegraphics[height=6cm, width=5cm, angle=0, scale=1.75]{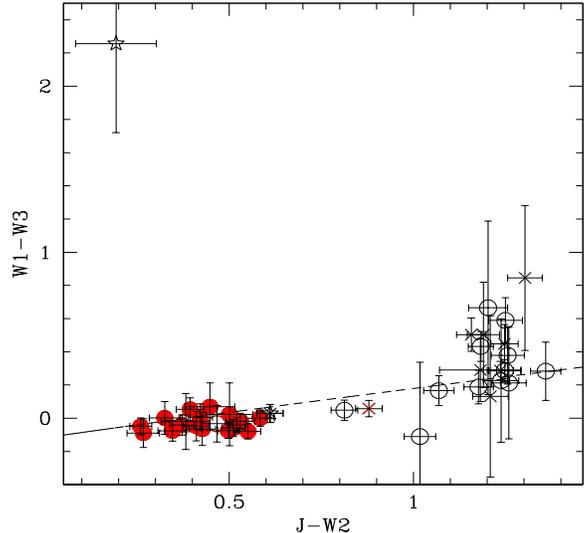}
\caption{ The solid  points show the J-W2 and W1-W3 colors for the brighter members of
the 16 systems in Table~\ref{TabG}. The open points are for the fainter star in each encounter. The
stars represent the white dwarfs in Table~\ref{TabM} and the crosses are those stars undergoing close
encounters with white dwarfs.
Many of these stars are not detected in the W4 bandpass, but those that do have a non-zero detection
in that bandpass are plotted in red, while the rest are plotted in black. The dashed line indicates the
black-body trend in these colours, as a function of effective temperature.
 \label{GIR}}
\end{figure}

In order to search for infrared excesses we cross-correlated the positions of the stars in Table~\ref{TabG} with
point sources detected in the 2MASS \citep{2MASS} and ALLWISE \citep{WISE} catalogs. Figure~\ref{GIR} shows the resulting set
of J-W2 versus W1-W3 colors for sources detected at 12$\mu$m and shorter wavelengths. There is no evidence of any sort of
infrared excess around any of the stars -- either the putative origin systems or the destinations.

In Figure~\ref{GIR} we also show the colours for the pairs featuring white dwarfs. Most white dwarfs derive
from stars originally in the same mass range as those in Figure~\ref{TabG}, and so may provide another
source for migration if civilisations survived the evolution of their host stars. Most of the white dwarfs are so
faint in the infra-red that they are not detected in WISE, but most of the other members of the encounter
pair are detected. The only source with potentially interesting colours is GD~84, which has a large
W1-W3 colour. However, the source is still faint at WISE wavelengths and the uncertainty in the W3
magnitude may be underestimated because of the uncertainty in the background level in this image. 

We have also examined the 2MASS and WISE magnitudes for the sample of lower mass stars in
Table~\ref{TabM}. These stars are mostly much fainter and, although there are a handful that
show anomalous WISE colours, all excesses can be explained by either source confusion or uncertainties
in the background levels at W3 and W4 wavelengths.

\begin{figure}
\includegraphics[height=6cm, width=5cm, angle=0, scale=1.75]{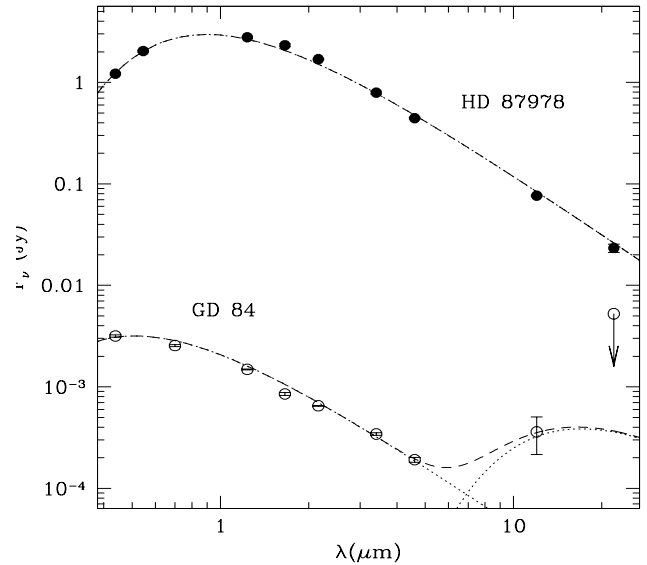}
\caption{The solid points represent the available photometry (from B to W4) for
the star  HD~87978 and the dashed line is a blackbody spectrum. The excellent fit indicates
no evidence for any significant infrared excess at long wavelengths. The open circles indicate
the available photometry for the white dwarf GD~84. In this case, there does appear to be an
excess at 12$\mu$m (W3 band) and we can explain this with a population of dust at a temperature
of 300~K (as indicated by the second dotted line). However, inspection of the WISE images
suggests that the background levels are quite uncertain at this magnitude and so this detection should be considered
very tentative.
 \label{Excess}}
\end{figure}

To assess what sort of limit these observations place on the presence of dust, Figure~\ref{Excess} shows the spectral energy distributions
for two stars -- GD~84 (the one star with a marginal detection of an excess) and  HD~87978 (our best case target in the sense
that it is a slightly evolved G star and so best corresponds to the scenario outlined in HZ21). In each case, we have taken the
WISE and 2MASS data, combined it with optical photometry from the literature, and fit with a black body of appropriate
temperature. We have then added, to the GD~84 data, a black body at temperature 300~K. This is based on the assumption
that any technological activity would occur at approximately habitable temperatures.

The measured flux can tell us the surface area that  a population of optically thin dust  is required to match. For a 
white dwarf of temperature $10^4$K and radius $10^9$cm, surrounded by a population of $10 \mu$m radius dust particles,
we require $N \sim 2 \times 10^{28}$ particles to intercept enough light to reradiate the observed flux in the infrared. Assuming
a dust density of 3g.cm$^{-3}$, this means a mass of $\sim 2 \times 10^{20}$g, which is very small --  comparable to an average 
Solar system asteroid. A similar calculation yields
a mass in dust $\sim 5 \times 10^{21}$g to match the observed flux of  HD~87978 at 24$\mu$m. This represents an upper limit
as there is no evidence of excess in the  HD~87978 flux. 

 Infrared excesses are a common feature of young planetary systems, as the dust from the late, collisional, stage of planetary assembly.
Some fragments must survive to late stages, because there is evidence for surviving rocky material in orbit around a significant fraction
of white dwarfs \citep{ZMKK}. Thus, we anticipate the rocky material from a surviving asteroid belt may be available for the construction of 
interstellar transport vehicles.
 Any estimate of the required mass is wildly speculative given our current lack of knowledge, but
we can set a baseline by considering what might be required for transportation of humans in such an episode. Given the expense
and danger involved, we can imagine that  construction would opt for extremes of efficiency. If the spaceships were constructed by
hollowing out asteroids (less expensive than boosting material from a planetary surface to space) we anticipate the smallest volume
per capita that is feasible. A contender for the highest population density long-term settlement is the Kowloon Walled City slum
of Hong Kong \citep{Kow}, with a population density of $\sim$ 1 million people per square km. The walled city was also 15 storeys
high, or about 60m, so that the volume associated with 1 million people is $\sim 6 \times 10^7 m^3$, or a volume of about 60 $m^3$
per person. The excavation of such a volume of rock implies a mass $\sim 2 \times 10^8$g per capita. If this were all converted to
dust, it would still allow for the construction of habitats for $\sim 10^{12}$ people. This does not allow for volumes for storage, or
engines or many other necessary items, but it illustrates that an alien construction need not produce an enormous infrared 
signature. 
 The only substantive conclusion we can draw is that we have no
significant evidence for anything unusual observed in the infra-red output of this sample of stars.

\subsection{Astrophysical Effects}

 We have also noted the possibility that close stellar encounters might induce comet showers.
The extent of the Solar Oort cloud is believed to extend well beyond the $10^4$~AU  threshold we have adopted, so it is 
natural to expect the stellar encounters identified here to affect any equivalent Oort clouds around these stars, should they
exist. However, the observable signals to be expected from such a comet shower require that gas or dust be released from the
comet to yield infrared excesses or transient spectral features in absorption, and this requires the comets to get close enough
to the star to begin to sublimate, which essentially requires getting within 5 AU \citep{Whip50}, in the case of a Sun-like star. 
Such a requirement imposes a delay $\sim 1 \rm Myr$ between the actual encounter and the peak of any expected signal \citep{Hut87,Dyb02}
-- corresponding roughly to the orbital timescale at the original location of the scattered comets. Thus, it is likely far too
early to expect much of a signal in the case of the sample of stars shown here.

Nevertheless, we can still estimate the size of the effect. Starting with a Sun-like Oort cloud, we take a total
of $10^{12}$ comets distributed uniformly within $5 \times 10^4$~AU of the host star. Several authors have made
more detailed models of the actual Solar Oort cloud, but this simple model captures the basic population demographics (e.g. \cite{Weiss96}).
In such a model, the local density of comets is $n \sim 2 \times 10^{-3}  \rm AU^{-3}$. If we ask how close to the perturbing star,
of mass $M_{pert}$, does a comet need to get to be strongly scattered, we can determine this by calculating the $\Delta V_{\perp}$ due to a scattering encounter (e.g. from \cite{BT87}) and set it to be comparable to the orbital velocity of the comet, at a distance $a$ from the host star of mass $M_*$. This yields an impact parameter (relative to the scatterer) of
\begin{equation}
b \sim 36 AU \left( \frac{M_{pert}}{0.3 M_{\odot}} \right) \left( \frac{1 \rm M_{\odot}}{M_*} \right)^{1/2} \left( \frac{50 \rm km.s^{-1}}{V_0} \right)\left( \frac{a}{10^4  \rm AU} \right)^{1/2}
\end{equation}
In passing through the cloud, the interloper perturbs a fraction $\sim 10^{-5}$ of the comets, assuming $M_{pert}=1 \rm M_{\odot}$, which is comparable to more detailed estimates (e.g. \cite{Dyb02}). This is not likely to be observable amongst the sample identified here -- because of the delay between perturbation and post-perturbation perihelion -- but it suggests
that the space density of stars experiencing comet showers should be similar to the density of stellar perturbations identified here.

One hope for  an immediate signal is for the perturber itself to cause sublimation in comets that pass close to it during its
passage through the Oort cloud. This has not been estimated in the above papers, so let us estimate the size of the effect within the context of our simple comet cloud model 
outlined above.
If we adopt the Whipple threshold of $\sim 5$~AU for the sublimation of water in Solar system comets, this amounts to an equivalent temperature of $T_{eq}=150$K on the sunward facing side of the comet. This implies the perturber must pass within
\begin{equation}
b_{sub} = 0.6 AU \left( \frac{R_{pert}}{0.3 \rm R_{\odot}} \right) \left( \frac{T_{pert}}{3550 \rm K}\right)^{2} \left( \frac{T_{eq}}{150 \rm K} \right)^{-2}
\end{equation}
to generate gas and dust by sublimation. Consideration of gravitational focussing increases this estimate by only 10 per cent.
 Therefore, for M dwarf perturbers, there is rarely a comet within a sufficiently large volume to generate any mass loss. In the case of a G-star perturber, the volume gets larger, and there is $\sim 1$ comet within the sublimation radius at any given time. However, the odds of this object passing across the line of sight of the observer is small. Over the course of a month, the line of sight intersecting a solar mass star will sweep out a volume $\sim 0.04 \rm AU^3$ (travelling at 50 km.s$^{-1}$), which is small compared to the volume $\sim 500 \, \rm AU^3$ per comet. 

Therefore, we conclude that the odds are negligible of seeing a transient absorption in one of these systems due to the perturber directly sublimating a comet in the Oort cloud.

\section{Conclusions}
\label{Conc}

HZ21 argue that the difficulties involved in migration between stars is minimized during close stellar encounters, and that the possibility of observing the operation
of extraterrestrial technologies may be enhanced during such episodes. As a result, in this paper we have searched the GAIA database of stars within 100~pc in order to
identify those unbound stellar pairs that  are likely to have close encounter in the recent past or near future.
. We identify a total of 132 pairs with  either a better than 1~per cent chance  of passing within $10^4$AU of one another, or with a mean value $\langle  1/r_{min}^2 \rangle < 0.1$.
Of these, 16 pairs feature a star of spectral type between K1--F3. These approximately Sun-like stars are the ones that correspond
most directly to our original hypothesis. Many of these pairs have measured radial velocities,  and we have used this information, where available, to improve
our estimates of the interaction metrics.

The motivation to undertake  interstellar migration may also be reduced if the star has a bound companion, as such a body would be much easier to reach. Examination of
our preferred sample for indications of binary companions leaves only  8/16 of the primaries in the most constrained sample as having no evidence for binarity. As a result, our preferred candidate
systems for observing signs of interstellar migration are HD~87978, HD~92577, HD~50669, HD~44006, HD~80790,  LSPM~J2126+5338, LSPM~J0646+1829 and  HD~192486.  The highest probability candidates amongst known planet hosts are the stars GJ~433 and HR~858.
 We suggest that searches for extraterrestrial technology include these stars in their search, as they provide a rationale for a hypothesis-driven search instead of the traditional blind search and thus may motivate deeper observations. Indeed, the
full sample of 16 systems in Table~\ref{TabG} are worthy of studying in this regard. In appendix~A we provide finding charts for each of these pairs, as well as for a few additional systems of potential interest (such as the most extreme members of the lower mass sample in Table~\ref{TabM}).

As an example of  such a search, we have searched the 2MASS and WISE databases for the stars in Table~\ref{TabG} to look for the possibility of infrared excesses due to
 large-scale astroconstruction in these systems. We find no evidence for an infrared excess in any of the G star systems. We also examined the encounters in the expanded sample (Table~\ref{TabM}) that involved white dwarfs -- since most of these stars are evolved from G or F main sequence stars. We find one marginal detection of a W3 excess around GD~84, but this must be regarded as tentative given the uncertainties in the background level at these wavelengths. From these non-detections we can constrain the mass in 10$\mu$m dust around these stars to be $<10^{21}$g.

Our study bears obvious resemblance to studies searching for the stars that will make close encounters with the Sun \citep{BJR16,BB21}. Those studies focus on stars close to the Sun, whose parallaxes are  often more accurate than those used here, and for which a much larger fraction have radial velocities available. In our case, we are limited primarily by the accuracy of the parallax and the unknown line-of-sight velocity, which renders our sample statistical in nature. It is likely that only one or two of the candidates in Table~\ref{TabG} will actually experience
an encounter within $10^4$~AU, but the benefits of having a well-defined, manageable sample compensate for this.


\begin{acknowledgements}

The author thanks the anonymous referee and  Ben Zuckerman for advice and comments on the manuscript. 
  This research was supported in part by grants to UCLA from NASA. This research has made use of NASA's Astrophysics Data System and the results of the RECONS consortium, which can be found at  www.recons.org.
  This work has made use of data from the European Space Agency (ESA) mission
{\it Gaia} (\url{https://www.cosmos.esa.int/gaia}), processed by the {\it Gaia}
Data Processing and Analysis Consortium (DPAC,
\url{https://www.cosmos.esa.int/web/gaia/dpac/consortium}). Funding for the DPAC
has been provided by national institutions, in particular the institutions
participating in the {\it Gaia} Multilateral Agreement.
This research has made use of the NASA/IPAC Infrared Science Archive, which is funded by the National Aeronautics and Space Administration and operated by the California Institute of Technology. This publication makes use of data products from the Two Micron All Sky Survey, which is a joint project of the University of Massachusetts and the Infrared Processing and Analysis Center/California Institute of Technology, funded by the National Aeronautics and Space Administration and the National Science Foundation. The Digitized Sky Survey was produced at the Space Telescope Science Institute under U.S. Government grant NAG W-2166. The images of these surveys are based on photographic data obtained using the Oschin Schmidt Telescope on Palomar Mountain and the UK Schmidt Telescope. The plates were processed into the present compressed digital form with the permission of these institutions.
This research has made use of the NASA Exoplanet Archive, which is operated by the California Institute of Technology, under contract with the National Aeronautics and Space Administration under the Exoplanet Exploration Program. 

\end{acknowledgements}


\newpage

\appendix

\section{Finding Charts}
\label{Finding}

Here we present the finding charts for the  16 systems in Table~\ref{TabG}, along with a handful of additional, potentially interesting, systems. We
show the relative proper motions of each of the members of the close encounter pair, to illustrate the nature of each encounter. Note that the magnitudes of the vectors are proportional
to the relative amplitude of the proper motions in each chart, but not between charts (i.e. there is no uniform scaling in proper motion that has been
applied across all charts). In the cases where radial velocities are measured for these stars, their direction is indicated as moving towards ($\bigodot$)
or away from ($\bigotimes$) the observer. 
 All images are taken from the ALADIN archive \citep{Aladin} of DSS2 images, except where specifically indicated.

\subsection{Encounters featuring Sun-like Stars}

\begin{figure}
\includegraphics[height=5cm, width=5cm, angle=0, scale=1.75]{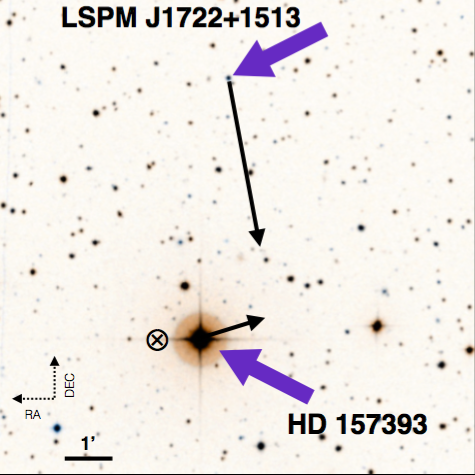}
\includegraphics[height=5cm, width=5cm, angle=0, scale=1.75]{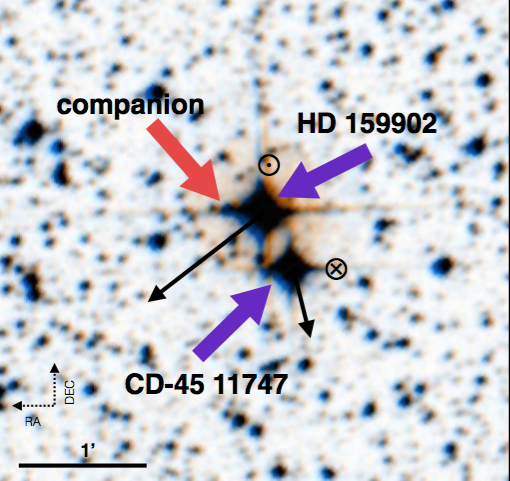}
\caption{The left panel shows the field surrounding the interacting pair HD~157393 and LSPM J1722+1513. The purple arrows indicate the
two stars involved in the interaction, while the black arrows indicate the proper motions of the two stars involved. The field is $10' \times 10'$ and the
co-ordinates of HD~157393 are  17$^{\rm h}$22$^{\rm m}$24$^{\rm s}$.9726448248, +15$^{\circ}$08$'$12$''$.868453175 (epoch J2000).
HD157393 is suspected to be a binary, based on the proper motion anomaly between Hipparcos and GAIA \citep{Kervella}. The right hand panel
shows the field surrounding the pair HD~159902 and CD-45 11747. The field is a little smaller here ($4' \times 4'$) because this pair is much closer
together. The position of HD~159902 is 17$^{\rm h}$39$^{\rm m}$27$^{\rm s}$.8792739684, -45$^{\circ}$53$'$10$''$.985009447 at epoch J2000. The red arrow indicates the probable proper motion companion to HD~159902.
 \label{Chart_157393}}
\end{figure}

\begin{figure}
\includegraphics[height=5cm, width=5cm, angle=0, scale=1.75]{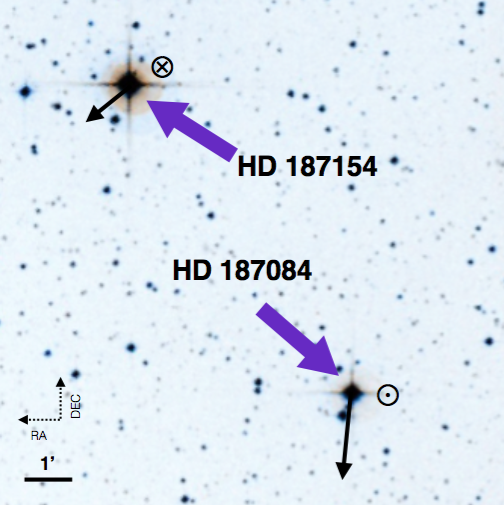}
\includegraphics[height=5cm, width=5cm, angle=0, scale=1.75]{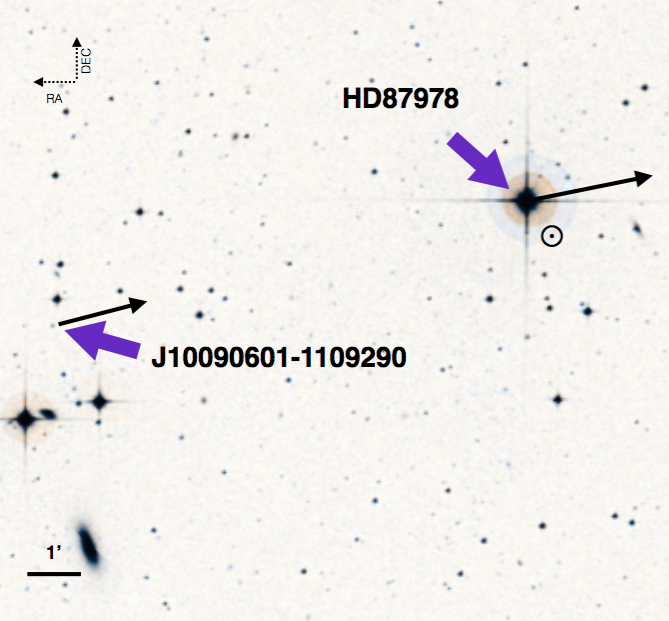}
\caption{The left panel shows the field surrounding the interacting pair involving the G1 star HD~187154 and the K1 star HD 187084. The purple arrows indicate the
two stars involved in the interaction, while the black arrows indicate the proper motions of the two stars involved. The field is $11' \times 11'$.  The position
of HD~187154 is 19$^{\rm h}$49$^{\rm m}$44$^{\rm s}$.0761820458, -32$^{\circ}$45$'$50$''$.287392501 at epoch J2000.
HD187154 is suspected to be a binary, based on the proper motion anomaly between Hipparcos and GAIA \citep{Kervella}.
 The right hand panel shows the field surrounding the G6IV star HD~87978 and the M dwarf J10090601-1109290. The field is $12.7' \times 12.7'$. The star
 HD~87978 is a potentially single Sun-like star, nearing the end of its main sequence lifetime. It's position is 10$^{\rm h}$08$^{\rm m}$26$^{\rm s}$.5768781632, -11$^{\circ}$06$'$54$''$.743846694 at epoch J2000.
 \label{Chart_187154}}
\end{figure}

\begin{figure}
\includegraphics[height=5cm, width=5cm, angle=0, scale=1.75]{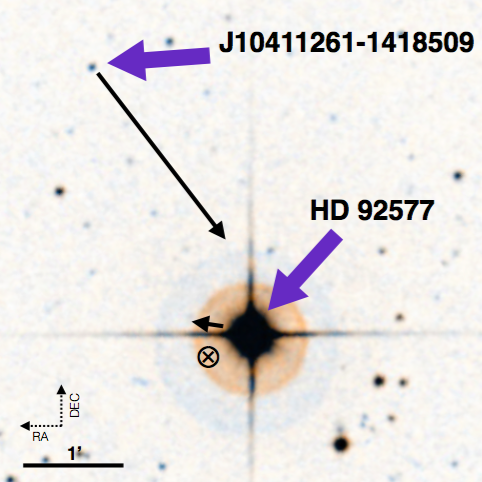}
\includegraphics[height=5cm, width=5cm, angle=0, scale=1.75]{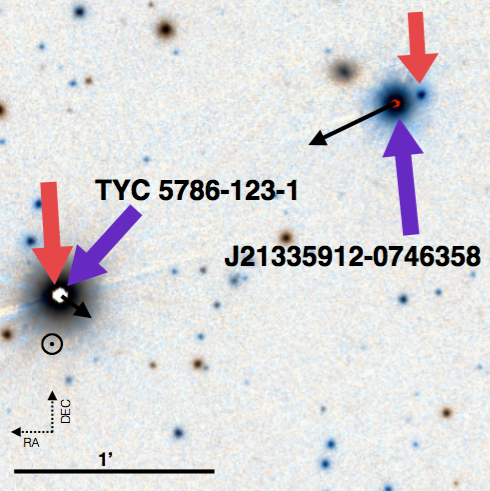}
\caption{ The left panel shows the field surrounding the pair involving the F2 star HD~92577 and the M dwarf J104111261-1418509. The field is a $5' \times 5'$.
The star HD~92577 is one of the potentially single Sun-like stars  in our table and its position is 10$^{\rm h}$41$^{\rm m}$06$^{\rm s}$.1151098272 ,-14$^{\circ}$21$'$29$''$.265485777 at epoch J2000.
The right panel shows the field surrounding the interacting pair  TYC~5786-123-1 and J21335912-0746358. The purple arrows indicate the
two stars involved in the interaction, while the black arrows indicate the proper motions of the two stars involved. The field is $2.5' \times 2.5'$ and was taken from the PanSTARRs survey because of the better angular resolution. The position of TYC~5786-123-1 is 21$^{\rm h}$34$^{\rm m}$06$^{\rm s}$.3864435104, -07$^{\circ}$47$'$35$''$.460488485 at epoch J2000.
Both stars involved in this encounter have common proper motion companions in the GAIA data, whose positions are marked by red arrows.
  \label{Chart_5786}}
\end{figure}

\begin{figure}
\includegraphics[height=5cm, width=5cm, angle=0, scale=1.75]{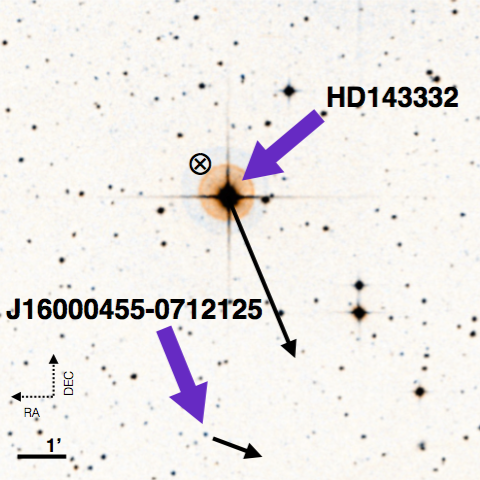}
\includegraphics[height=5cm, width=5cm, angle=0, scale=1.75]{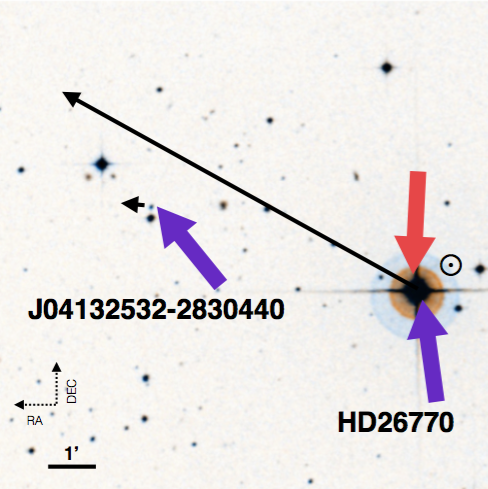}
\caption{The left panel shows the 
  encounter between the F5 star HD~143332 and the M star J16000455-0712125. The field is $10'\times 10'$.
 The position of HD~143332 is 16$^{\rm h}$00$^{\rm m}$02$^{\rm s}$.6498506984, -07$^{\circ}$07$'$21$''$.680091901 at epoch J2000.
The right hand panel shows the encounter between the binary star HD~26770 and the M dwarf J04132532-2830440.  The red arrow indicates the presence of a binary companion. The field is $10' \times 10'$. The position of HD~26770 is 04$^{\rm h}$13$^{\rm m}$00$^{\rm s}$.52912, -28$^{\circ}$32$'$26$''$.2934 at epoch J2000.
  \label{Chart_BD+00}}
\end{figure}

Figure~\ref{Chart_157393} shows the finding charts for the first two entries in Table~\ref{TabG}. The left hand panel shows the encounter involving the F8 star
HD~157393. In this case, the closest approach is estimated to lie in the future, and the proper motion vectors appear to be converging. 
In the right hand panel,
we see the encounter involving the G5 star HD~159902. The pair, in this case, is much closer and the closest approach is estimated to have occurred in the recent
past. Thus, the proper motions now appear to be diverging. The star HD~159902 also appears to have a wide ($a_{\perp}\sim 1000$~AU) proper motion companion in the GAIA data.

 Figure~\ref{Chart_187154} shows two more finding charts, for the star HD~187154 (left panel) and HD~87978 (right panel). The former case now shows diverging proper
motions and the closest approach is estimated to have occurred in the past. The latter case is currently showing converging proper motions and is estimated to undergo close
approach in the future. HD~87978 is also one of the stars in Table~\ref{TabG} for which there is, as yet, no evidence of a bound companion.

 The left panel of Figure~\ref{Chart_5786} shows the finding chart for the encounter featuring the F2V star HD~92577.
The right panel of Figure~\ref{Chart_5786} shows the finding chart for a binary-binary encounter. In this case, the minimum encounter distance is expected to be much wider than
either of the binary separations ($a_{\perp}=712$~AU for TYC-5786-123-1 and $a_{\perp}=310$AU for the M dwarf pair). 

The left hand panel of Figure~\ref{Chart_BD+00}  shows the encounter of HD~143332 with another M dwarf. This is projected to reach closest approach $\sim 10^4$ years in the future, consistent with the
evidently converging proper motions. HD~143332 is also suspected to be a binary \citep{Kervella}.  The right hand panel shows the encounter of the G0V star HD~26770 with an M dwarf. The HD~26770 system is actually a wide binary containing two G stars.

\begin{figure}
\includegraphics[height=5cm, width=5cm, angle=0, scale=1.75]{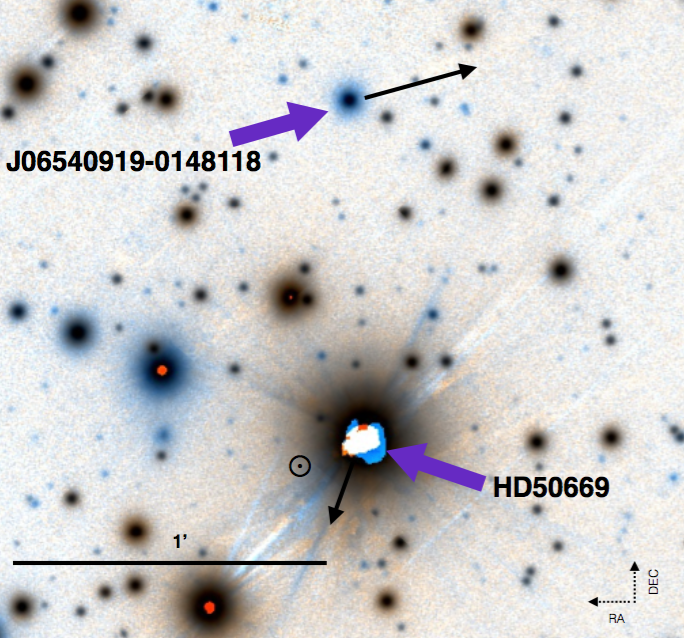}
\includegraphics[height=5cm, width=5cm, angle=0, scale=1.75]{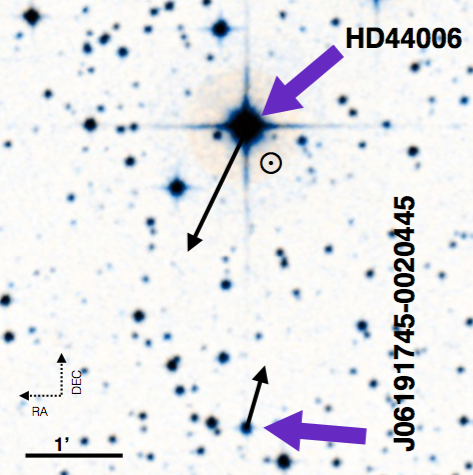}
\caption{The left panel shows the  pair involving the G8IV star HD~50669 and the M star J06540919-0148118. The purple arrows indicate the
two stars, while the black arrows indicate the proper motions. The field is $2.2' \times 2.2'$, and is drawn from the PanSTARRS survey.
The position of HD~50669 is 06$^{\rm h}$54$^{\rm m}$08$^{\rm s}$.8776552042, -01$^{\circ}$49$'$20$''$.549547800 at epoch J2000.
 The right hand panel shows the encounter between the F5 star HD~143332 and the M star J16000455-0712125. The field is $10'\times 10'$.
 The position of HD~143332 is 16$^{\rm h}$00$^{\rm m}$02$^{\rm s}$.6498506984, -07$^{\circ}$07$'$21$''$.680091901 at epoch J2000.
  \label{Chart_50669}}
\end{figure}

\begin{figure}
\includegraphics[height=5cm, width=5cm, angle=0, scale=1.75]{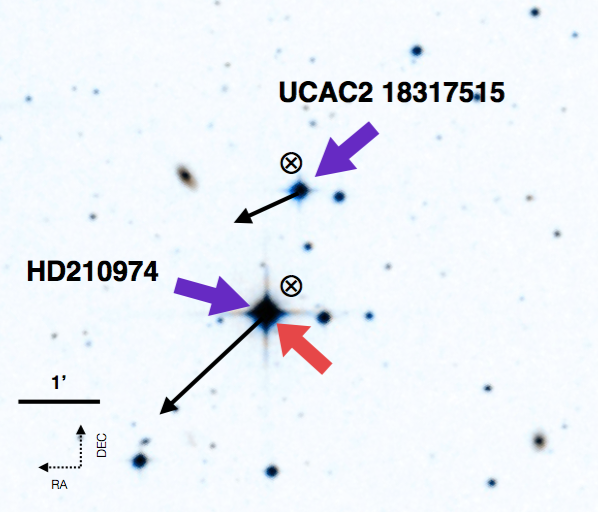}
\includegraphics[height=5cm, width=5cm, angle=0, scale=1.75]{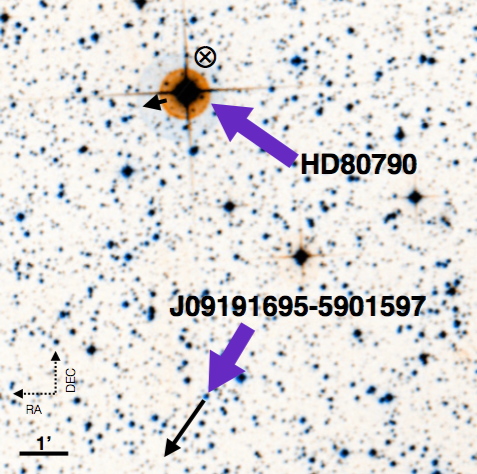}
\caption{The left panel shows the encounter between the binary star {HD210974} and the star UCAC2~18317515. The purple arrows indicate the
two stars, while the black arrows indicate the proper motions. The field is $7.5' \times 7.5'$. 
The position of  HD~210974 is 22$^{\rm h}$14$^{\rm m}$38$^{\rm s}$.26886, -32$^{\circ}$12$'$23$''$.2520  at J2000.
The right hand panel shows the encounter between the F6 star HD~80790 and the M star J09191695-5901597. The field is $10'\times 10'$. The 
position of HD~80890 is 09$^{\rm h}$19$^{\rm m}$19$^{\rm s}$.9662065270, -58$^{\circ}$55$'$45$''$.646644916 at epoch J2000.
  \label{Chart_210974}}
\end{figure}

The left hand panel of Figure~\ref{Chart_50669} shows the interaction of the  G8IV star HD~50669 with an M star. This is an apparently single star approaching the end of its main sequence lifetime and thus another prime candidate for an interstellar migration event such as those postulated in HZ21. 
The  right hand panel of Figure~\ref{Chart_50669} shows the finding chart for the star HD~44006, another isolated G star undergoing an encounter with an M dwarf.

 The left hand panel of Figure~\ref{Chart_210974} shows an encounter between another binary featuring G8 and K0V stars, with a fainter star UCAC2~18317515. This system
likely went through its closest approach fairly recently. The right hand panel shows 
the field around the HD~80790. This star underwent a close passage with the M star J09191695-5901597 a few thousand years ago and is now diverging. The star HD~80790 is identified as a binary in the literature.

\begin{figure}
\includegraphics[height=5cm, width=5cm, angle=0, scale=1.75]{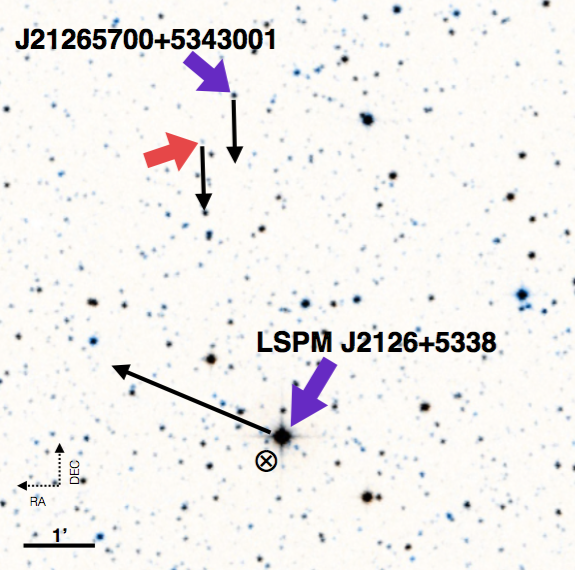}
\includegraphics[height=5cm, width=5cm, angle=0, scale=1.75]{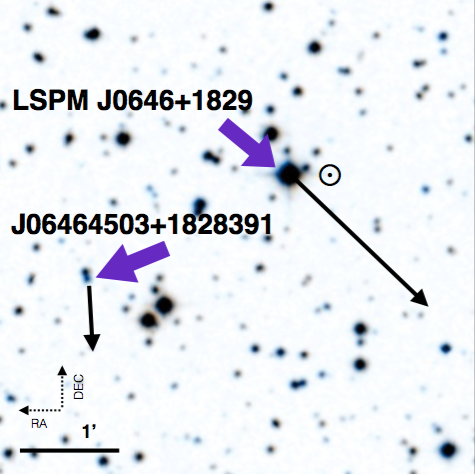}
\caption{The left panel shows the field containing the star LSPM J2126+5338 and the M star J21265700+5343001. The purple arrows indicate the
two stars, while the black arrows indicate the proper motions. The field is $8' \times 8'$.  The
position of LSPM J2126+5338 is 21$^{\rm h}$26$^{\rm m}$52$^{\rm s}$.8508484028, +53$^{\circ}$38$'$22$''$.323798302 at epoch J2000.
The right hand panel shows the  pair of stars LSPM~J0646+1829 and  J06464503+1828391. The field is $5'\times 5'$.
The position of LSPM~J0646+1829 is 06$^{\rm h}$46$^{\rm m}$36$^{\rm s}$.3522038800, +18$^{\circ}$29$'$43$''$.045073347 at epoch J2000.
  \label{Chart_2126}}
\end{figure}

The left hand panel of Figure~\ref{Chart_2126} shows the field of the encounter featuring the star LSPM J2126+5338 with an M dwarf binary. This is actually the highest probability encounter in Table~\ref{TabG}. Finally, the right hand panel shows the encounter featuring the star LSPM~J0646+1829. This is another of the single Sun-like stars undergoing a close encounter. 

\begin{figure}
\includegraphics[height=5cm, width=5cm, angle=0, scale=1.75]{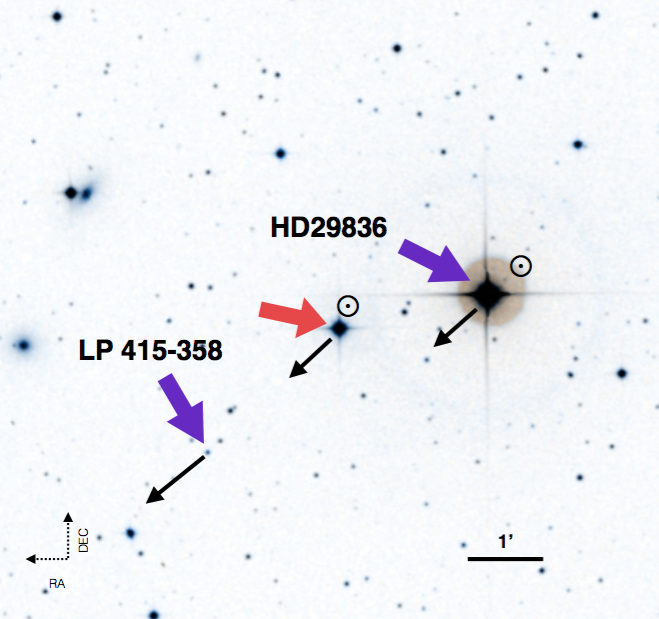}
\includegraphics[height=5cm, width=5cm, angle=0, scale=1.75]{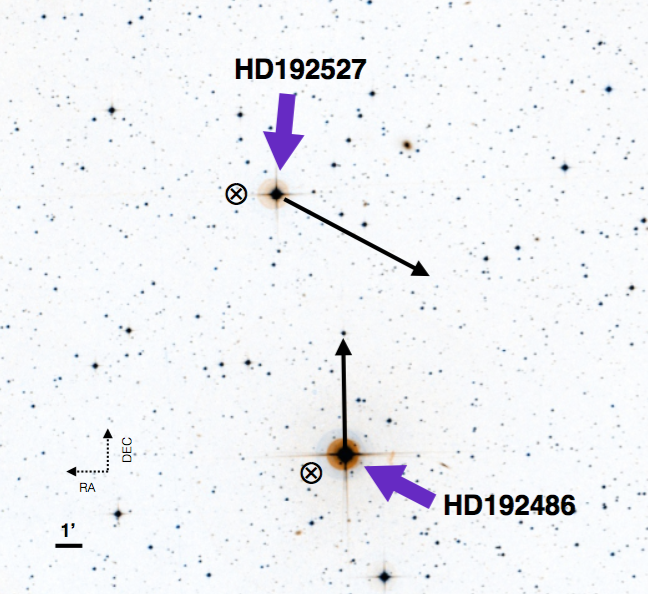}
\caption{ The left panel shows the encounter of the wide proper motion pair HD~29836 and HD~28570 (indicated with the red arrow) with the M dwarf LP~415-358. 
The principals in the encounter are indicated by purple arrows and the proper motions by the black arrows. The field
shown here is $8.8' \times 8.8'$. The position of HD~29836 is 04$^{\rm h}$59$^{\rm m}$37$^{\rm s}$.3055400488, +83$^{\circ}$11$'$46$''$.005522306 at epoch J2000.
The right hand panel shows the encounter between HD~192486 and HD~192527. The field size here is $24.5' \times 24.5'$. The position of HD~192486 is
 20$^{\rm h}$16$^{\rm m}$26$^{\rm s}$.4481841445, -35$^{\circ}$11$'$54$''$.476960125 at epoch J2000.
  \label{Chart_29836}}
\end{figure}

 The final figure in this section contains the final two entries in Table~\ref{TabG}. Neither of these encounters has a minimum likely passage within
$10^4$AU, but both have $<(1/r_{min})^2>$ above the threshold value of 0.1. The left hand panel shows the encounter of a GV star HD~29836,
with an M dwarf LP~415-358. HD~29836 has a wide K7 proper motion companion, HD~285970. Despite the low values of $P_{\mu}$ and $P_{rv}$, this
encounter has the equal fifth highest value of  $\langle 1/r_{min}^2\rangle$ in Table~\ref{TabG}. The right hand panel shows the encounter of an F2V star
HD~192486 and a G5V star HD~192527. The two stars share a common radial velocity, but clearly discrepant proper motions. This pair lies close to the
lower limit of our $\Delta v$ cutoff, but there is no indication of excess astrometric noise that would suggest one of the sources is a partially resolved binary
whose motion may have distorted the relative proper motions.

\subsection{Other Systems of Interest}

\begin{figure}
\includegraphics[height=5cm, width=5cm, angle=0, scale=1.75]{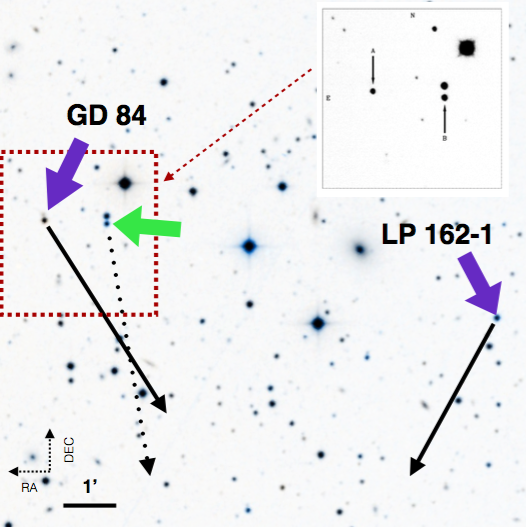}
\includegraphics[height=5cm, width=5cm, angle=0, scale=1.75]{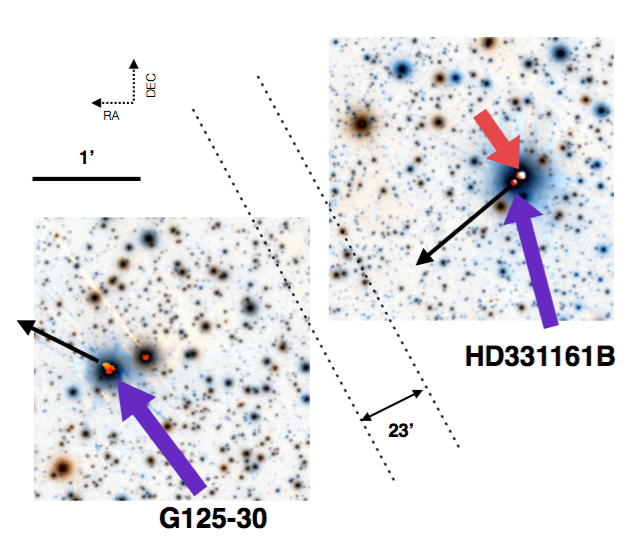}
\caption{The left panel shows the  interacting pair involving the white dwarf  GD~84 and the M star LP~162-1. The purple arrows indicate the
two stars, while the black arrows indicate the proper motions. The green arrow in this field indicates the star initially
identified as a possible companion by \cite{FBZ} but which now lies definitively farther away. The inset at the upper right shows the finding chart from \cite{Farihi_thesis}, which identifies the putative companion as star~B. The red dashed square indicates the corresponding field of view in our image.
The dotted arrow shows the proper motion for this star as well. 
 The field is $5' \times 5'$. The position of GD~84 is 07$^{\rm h}$18$^{\rm m}$01$^{\rm s}$.9005268754, +45$^{\circ}$47$'$53$''$.06535215 at epoch J2000.
The right hand panel shows the encounter between the M dwarfs HD~331161B and the M star G125-30. Here we show only postage stamp images around each star, because these
stars are separated by 23 arcminutes on the sky. Of all the pairs in our sample, this is the closest to the Sun. The position of HD~331161B is 19$^{\rm h}$46$^{\rm m}$24$^{\rm s}$.2211145800, +32$^{\circ}$00$'$57$''$.783307309 at J2000. The fields here are taken from the PanSTARRS survey.
  \label{Chart_GD84}}
\end{figure}

The encounter involving the white dwarf GD~84 features a marginal detection of an infra-red excess. Figure~\ref{Chart_GD84} shows this encounter and also indicates the
star tentatively identified by \cite{FBZ} as a companion -- see Figure 4.4 of \cite{Farihi_thesis} for the original finding chart. The  identification was tentative because the stellar proper motion measurement was compromised by confusion with the nearby background star. GAIA measurements now reveal that this is not at the same distance nor does it share the proper motion of GD~84. 

The right hand panel of Figure~\ref{Chart_GD84} shows the identification of the two stars taking part in the encounter that is closest to the Sun -- that of HD~331161B and G125-30. This pair is so close to the Sun (14 pc away) that a full finding chart that spans the region between the pair includes so many stars as to be essentially unreadable. Thus, in this chart we
provide two zooms in on the stars in question, separated by $\sim 23'$ on the sky.

\begin{figure}
\includegraphics[height=5cm, width=5cm, angle=0, scale=1.75]{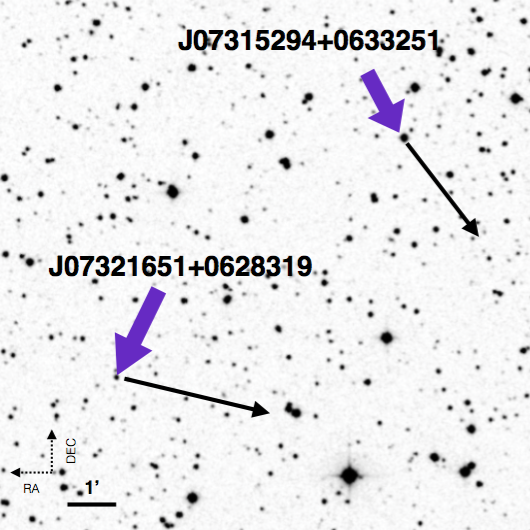}
\includegraphics[height=5cm, width=5cm, angle=0, scale=1.75]{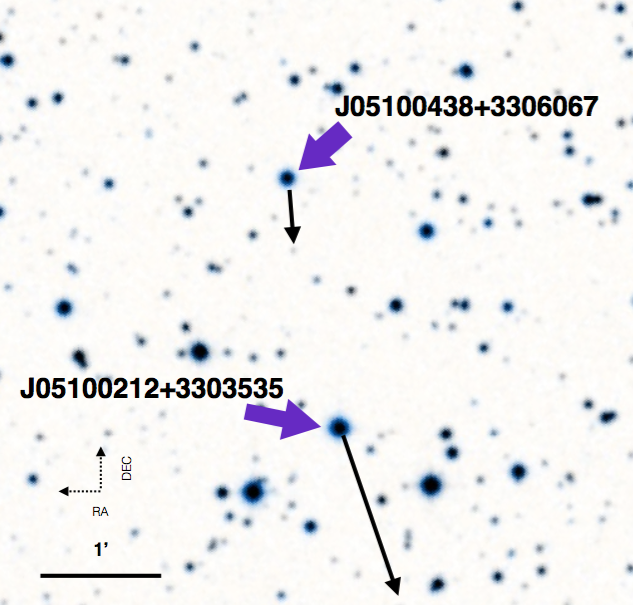}
\caption{The left hand panel shows the approach that has the smallest likely perihelion of our entire sample, with a possible minimum of only 105~AU, featuring two M dwarfs J07315294+0633251 and J07321651+0628319. The field of view here is $11' \times 11'$.
The right hand panel shows the pair that has the highest estimated probability of a passage within $10^4$AU, namely the pair J05100438+3306067 and J05100212+3303535.
The field of view here is $5.3'\times 5.3'$.
  \label{Chart_lowb}}
\end{figure}

 Figure~\ref{Chart_lowb} shows the encounters with the most extreme parameters. In the left panel, we see the interaction of the two M-dwarfs
J07315294+0633251 and J07321651+0628319, whose minimum likely close passage distance is estimated to be $\sim 105$~AU. The converging nature of the proper motions indicates that, with the right radial velocities, this can have a very close passage in $\sim 10^4$~years. The right hand panel shows the pair with the highest estimated probability of a close interaction. In this case, the high probability is not driven by the proper motions necessarily, but rather by the close correspondence in the distances to the two stars.

\begin{figure}
\includegraphics[height=5cm, width=5cm, angle=0, scale=1.75]{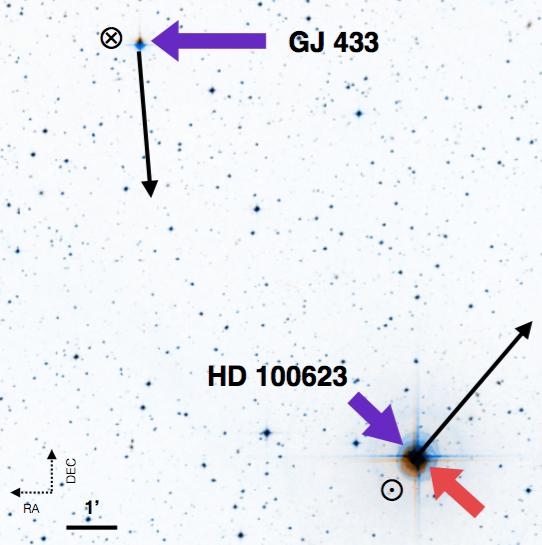}
\includegraphics[height=5cm, width=5cm, angle=0, scale=1.75]{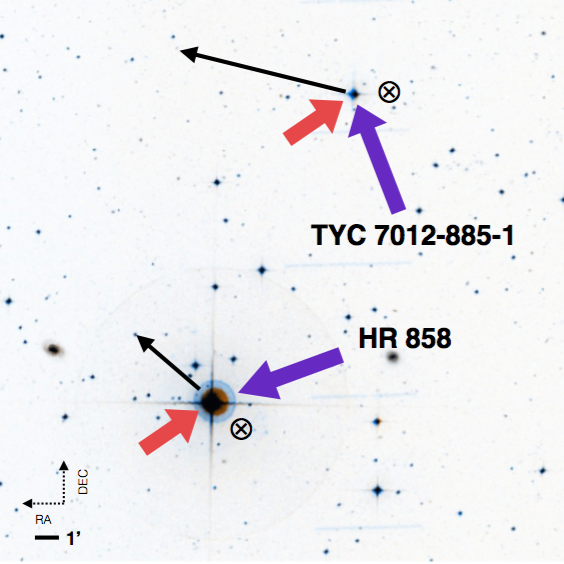}
\caption{The left hand panel shows the encounter featuring the planet host star GJ433, a $\sim 0.5 M_{\odot}$ M2V star that hosts two short period Super-Earths and a long period, Saturn-class planet. The field of view of this figure is $11' \times 11'$. The position of star GJ~433 is 11$^{\rm h}$35$^{\rm m}$26$^{\rm s}$.9474098036, -32$^{\circ}$32$'$23$''$.883014588 at epoch J2000.
The right hand panel shows the encounter featuring HR~858, a $\sim 1.2 M_{\odot}$, F6V star, which hosts three short-period Super-Earths discovered by TESS. This star is converging to an encounter with another binary pair in the near future. The field of view here is $24' \times 24'$. The position of HR~858 is 02$^{\rm h}$51$^{\rm m}$56$^{\rm s}$.2464213292 ,-30$^{\circ}$48$'$52$''$.259083285 at epoch J2000.
  \label{Chart_planets}}
\end{figure}

Our final finding charts , in Figure~\ref{Chart_planets}, show the two exoplanet host systems that have the closest encounters within the context of our analysis. The left hand panel shows the star GJ~433, which
hosts three planets, and its trajectory relative to a binary containing a K0V star and a white dwarf. The closest approach here likely happened a few thousand years ago and the pair is now diverging. On the right hand side of Figure~\ref{Chart_planets}, we show the encounter featuring HR~858. This system is likely converging towards its closest approach in the future. This encounter is once again an encounter between two binaries.

%




	\bibliographystyle{aasjournal}
	\bibliography{msBH_arxiv}




\end{document}